%% file: spetty_etgs_v7-pp2.tex
\documentstyle[natbib,preprint2,longabstract,11pt]{aastex}    

\input{ajbib.tex}

\input{figs-tables_spetty_etgs_v7-pp2.tex}

\def\M20{$M_{20}$}
\def\micron{$\mu$m}
\def\K{K$\rm_{s}$}
\def\zsun{Z$_\odot$}
\def\Rin{$R_{\rm in}$}
\def\Rout{$R_{\rm out}$}
\def\delt{$\Delta_T$}
\def\nuvr{NUV-$r$}
\def\nuvg{NUV-$g$}
\def\gr{$g$-$r$}
\def\gi{$g$-$i$}
\def\Teff{$T_{\rm eff}$}
\def\MgII{\ion{Mg}{2}}
\def\tage{$t_{\rm age}$}
\def\fbhb{$f_{\rm BHB}$}

\begin{document}

\title{UV-bright nearby early type galaxies observed in the mid-infrared: evidence for a multi-stage formation history by way of WISE and GALEX imaging}
\author{S.~M. Petty\altaffilmark{1,2}, J.D. Neill\altaffilmark{3}, T.~H. Jarrett\altaffilmark{4}, A.~W. Blain\altaffilmark{5}, D.~G. Farrah\altaffilmark{1}, R.~M. Rich\altaffilmark{2}, C.-W. Tsai\altaffilmark{4},  D. J. Benford\altaffilmark{6}, C. R. Bridge\altaffilmark{3}, S. E. Lake\altaffilmark{2}, F. J. Masci\altaffilmark{7}, E.~L. Wright\altaffilmark{2} }

  \altaffiltext{1}{Dept. of Physics, Virginia Tech Blacksburg, VA, 24061}
  \altaffiltext{2}{Dept. of Physics \& Astronomy, The University  of California, Los Angeles, CA 90095}
  \altaffiltext{3}{Div. of Physics, Math \& Astronomy, California  Institute of Technology, Pasadena, CA 91125}
   \altaffiltext{4}{Astronomy Department, University of Cape Town, Rondebosch, 7701, RSA}
   \altaffiltext{5}{Physics \& Astronomy, University of Leicester, Leicester,LE1 7RH, UK}
   \altaffiltext{6}{NASA, Goddard Space Flight Center, Greenbelt, MD 20771}
    \altaffiltext{7}{IPAC, California Institute of Technology, Pasadena, CA 91125}
    \newpage

 \begin{abstract}
In the local Universe, 10\% of massive elliptical galaxies are observed to exhibit a peculiar property: a substantial excess of ultraviolet emission (UVX) over what is expected from their old, red stellar populations. Several origins for the UVX have been proposed, including a population of hot young stars, or a population of old, blue horizontal branch or extended horizontal branch (BHB or EHB) stars that have undergone substantial mass loss from their outer atmospheres. We explore the radial distribution of ultraviolet excess (UVX) in a selection of 49 nearby E/S0-type galaxies by measuring the extended photometry in the UV through mid-IR with GALEX, SDSS and WISE. We compare UV/optical and UV/mid-IR colors with the \citet{con10} Flexible Stellar Population Synthesis models, which allow for the inclusion of EHB stars. We find that combined WISE mid-IR and GALEX UV colors are more effective in distinguishing models than optical colors, and that the UV/mid-IR combination is sensitive to EHB fraction. There are strong color gradients with the outer radii bluer than the inner half-light radii by $\sim1$ magnitude. This color difference is easily accounted for with a BHB fraction increase of 0.25 with radius. We estimated the average ages for the inner and outer radii are 7.0 $\pm 0.3$ Gyr, and 6.2 $\pm 0.2$ Gyr, respectively, with the implication that the outer regions are likely to have formed $\sim1$  Gyr after the inner regions. Additionally, we find that metallicity gradients are likely not a significant factor in the color difference. The separation of color between the inner and outer regions, which agrees with a specific stellar population difference (e.g., higher EHB populations), and the  $\sim 0.5$ to 2 Gyr age difference suggests multi-stage formation. Our results are best explained by inside-out formation: rapid star formation within the core at early epochs ($>4$ Gyr ago) and at least one later stage starburst event coinciding with $z\sim 1$. \\
\clearpage
\end{abstract}

\section{Introduction}

Early type galaxies (ETGs; ellipticals and lenticulars) are the oldest class of galaxy, and the end stage of galaxies formed at early epochs. Through multiple epochs they evolved into the most massive nearby galaxies, comprising at least 50\% of the local stellar mass \citep{bel03}. The most massive of these are older and more metal rich \citep{nel05}. Additionally, modified hierarchical simulations show that the most massive ellipticals have relatively shorter formation time-scales, more progenitors, and later assembly times \citep{deluc06}. Down-sizing scenarios \citep{cow96} predict that the stars formed at earlier epochs over a shorter time period. 

Since the majority of the luminosity for ETGs arises from the red part of the spectrum, they have been characterized as ``red and dead.'' Still, approximately 80\% of nearby ETGs are detected in the UV \citep[i.e., GALEX/NUV,][]{scha07} . The primary source of the UV emission is currently unclear. Stellar synthesis models have yet to converge on the combination of stellar properties that lead to a significant number of post main-sequence stars that are longterm producers of UV, creating a UV-excess (UVX). In ETGs, the UVX appears as a sudden increase in the flux in spectral energy distributions (SEDs) blueward of 2500 \AA, and was originally named the ``UV upturn'' \citep[e.g., ][]{cod79,oco92,bro95}. The origin of the UVX has been debated for several decades, with possible sources being recent star formation, metal-poor $>10$ Gyr-old horizontal branch (HB) stars, younger metal-rich HB stars, or a combination of these \citep[see][for a review]{oco99}.

Methods to discern the properties between red and dead ETGs versus galaxies with more recent star formation include: 1) photometric narrow- and wide-band properties, e.g., H$\alpha$, color-magnitude relations; 2) structural properties, such as detection of disk remnants or merger signatures; 3) spectroscopic detection of age-dependent lines, such as Mg, [OII], and H$\beta$. The  presence of recent star formation deduced through ionized gas emission in ETGs has been examined for decades. There is little question that some ETGs have signatures of recent star formation in the their nuclei and even disk remnants, but detecting this emission requires sensitive, high resolution imaging or spectroscopy, especially at the outer radii.

General consensus in the literature is that the source of UVX is likely generated from older, extreme stars ($>2$ Gyr) to create the temperatures needed based on the UV slope. However, evidence for recent star formation in ETGs has also been observed \citep[e.g.,][]{sal05,yi05,scha07,kav07b,kav09}. The presence of trace star formation in ETGs is generally correlated with environment, mostly detected in low-density environments \citep[e.g.,][and references therein]{sar07,sha10}, and is organized into disk-like and ring configurations.

There is no clear spectral distinction between the post-main-sequence or low-mass stellar populations that can produce a large amount of UV emission. Blue post-main-sequence stellar populations, such as blue horizontal branch stars (BHBs) with extreme temperatures and low-mass stellar envelopes, are promising candidates. At higher temperatures (\Teff$\gtrsim 14000$ K) BHBs are called extended (sometimes `extreme') HBs \citep[EHBs;][]{cod79, bur88, gre90,fer93,bro95,bro03, atl09, con10, yi11}. \citet{dor93} find that the driving factor to increase the temperatures of BHBs is the mass of the envelope in the zero-age HB phase, where a smaller envelope allows for higher temperatures; higher metallicities will result in a bimodal distribution of temperatures for envelope masses of 0.05-0.15 M$_{\odot}$. The morphology of the HB in the color-magnitude diagram is critically determined by the Reimers mass-loss parameter \citep[$\eta$; e.g.,][]{per11}, which describes the efficiency of the mass-loss of the outer envelope in the HB phase. The complexities of the BHB phase are not easily isolated with broad-band photometry or low-resolution spectroscopy. It is difficult to distinguish the BHB stars from the upper main sequence stars, using line strengths. These complications are in addition to the familiar age-metallicity degeneracy \citep{lee00,tra05, con10}. 

A metal-rich star would produce UVX through significant mass-loss in the red giant branch phase: a resulting lower mass HB (AGB-manqu\'e) star could sustain \Teff $\sim 20000$ K for $\sim 10$ Myr \citep[e.g.,][]{oco99}. Low-metallicity stars may also play a part, since the oldest populations are thought to be metal poor; however, models suggest that the UVX is not present until $>15$ Gyr for Z$\lesssim 0.5$ \zsun~\citep{yi98}. Observations show a lack of  a UV upturn in metal-poor galactic clusters as compared with giant ellipticals \citep[][and references therein]{yi98}. Additionally, models that include UV upturn show that solar, or higher metallicities allow UVX at younger ages ($<5$ Gyr), compared with metal-poor populations \citep[$>10$ Gyr; see Fig. 9 in ][]{yi98}. However, this is strongly dependent on the mass-loss efficiency. It would require a significant amount ($>20\%$ of HB stars) to produce the observed FUV in galaxies \citep{oco99}, making the ages and lifetime of EHB phases an important factor. In globular clusters, up to 30\% of the HB stars appear to be in an EHB phase \citep{kal92,lie94}.

Binary stars provide a promising alternative, since the environments can host the conditions for rapid loss of the hydrogen envelope, forming hot helium-burning stars. There are no metallicity requirements to produce UVX in binaries. Observed candidates for these binaries include hot subdwarf (sdBs) stars in globular clusters \citep[see review by][]{heb09}. \citet{han07} model the UV excess produced by binary sdBs, in which stellar populations as young as 1 Gyr can reach these extreme temperatures and reproduce the colors seen for ETGs.

How massive ellipticals evolve to their present state is important for understanding galaxy evolution. The theories need to integrate many factors, including the morphologies and age-metallicity differences observed between more and less massive ETGs. Mergers play an important role in explaining these factors. For example, dissipational (gas-rich) mergers would seed new star formation, while a dissipationless (gas-poor) merger would add mass via combining older stellar populations. An older stellar population in the outer halo may also be explained by outside-in cessation, where the star-formation ceases first in the disk \citep[e.g.,][]{pip04,pip06}. In an inside-out cessation, a core forms through very rapid starbursts at high redshifts, and through major multi-stage growth at later epochs, the outer regions form via minor mergers and accretion of hot gas from the immediate environment \citep[e.g.,][]{dad05,nel12,saracco12}.

The cores of nearby ETGs (within a half-light or effective radius) have been studied in much greater detail than the extended regions. The most recent and comprehensive study, the SAURON survey \citep{dez02}, uses the SAURON integral-field spectrograph to survey 72 nearby galaxies; 48 are ETGs. Among the sample, a kinematic separation is identified that separates the ETGs into subpopulations of slow and fast rotators based on their projected angular momentum \citep{ems07}. The slow rotators tend to be less compact and comprised of older stellar populations \citep[$>8$ Gyr;][]{mcder06}. Additionally, a slight trend with \MgII~ and FUV-NUV color is found for the slow-rotators. The fast rotators have ages of $<5$ Gyr. \citet{sha10} use mid-IR colors with {\it Spitzer} to trace low-level star formation through PAH emission in the sample. Most significantly, they find that the slow rotators have no sign of PAH emission and are quiescent over an $\sim8$ Gyr period.

In summary, much literature has been written to address the origin of UV emission in ETGs, and how they may have evolved into their present form. Resolved spectral analysis on ETGs has mostly focused on scales slightly larger than the effective radius, because of practicality. For example, \citet{jeo12} use resolved spectroscopy to test the Burstein relation on galaxies with extended UV. However, the Mg lines used are within the effective radii. Another example is the work of \citet{car11}, who compare the  UV and near-IR colors for the total flux and and central regions of ETGs, to determine if the FUV excess is related to velocity dispersion, metallicity or abundance ratio. However, their radial profiles do not extend beyond $\sim 30\arcsec$ from the galactocentric radius of each galaxy. These studies and similar ones do not use stellar population modelling to study the central and outer regions simultaneously to a spatial extend and resolution that is possible with photometry. In this paper, we conduct such an analysis as the first step in an expanded, and detailed, series of papers.

We use imaging from the recently completed Wide-field Infrared Survey Explorer (WISE) mid-IR imaging to explore the UV to mid-IR color-space of ETGs with UVX. The two main objectives in our study are to: 1) determine a how photometry between the UV and mid-IR can be used to interpret the likely stellar populations contributing to the UV emission; 2) use radial information from these results to interpret how these galaxies may have evolved. We do this over the total flux of the galaxies in our sample to analyze the color differences between the inner and outer regions. WISE provides the all-sky capabilities for a complete study. In \S \ref{selection}, we discuss our selection, and the detailed extended photometry. In \S \ref{ssptemp} we describe the stellar population synthesis models and parameters used for our analysis. We explore the color gradients, spatial distribution and possible source of the UVX in our sample in \S \ref{results}. We conclude with a discussion on the physical origin and the possible formation scenarios that may explain our results in \S \ref{discussion}. Where necessary, we adopt a $\Lambda$CDM cosmology, ($\rm \Omega_m, \, \Omega_\Lambda, \, H_0$) =(0.27, 0.73, 71 $\rm km \,s^{-1}\, Mpc^{-1}$).  

\section{Sample Selection}\label{selection}
We selected ETGs from the GALEX-Ultraviolet Atlas of Nearby Galaxies \citep{gil07}. This atlas includes galaxies from the GALEX Nearby Galaxy Survey and galaxies within that field and other fields with similar or greater depth. They select galaxies based on optical parameters from \citet{devau91} with $\rm \mu_B = 25 \, mag \, arcsec^{-2}$ isophote diameters larger than $1\arcmin$; they also use the \citet{devau91} galaxy types to label the morphologies. The atlas contains a total of 1034 galaxies where 893 have both FUV and NUV detections. For our sample, we selected by morphological type E or S0, resulting in a sample of 125 galaxies all at $z<0.06$. We then applied three further criteria. First, since we selected the sources during the execution of the WISE survey, not all sources had full-depth WISE data at the time the analysis was performed. We therefore removed the sources without full-depth WISE data. This amounted to a random cull of the parent sample. Second, we removed sources that did not have coverage in both GALEX channels. Third, we removed sources that lay near bright foreground objects, or had obvious artifacts in the data. This resulted in a final sample of 49 reliable objects, at  a mean redshift of 0.02. Since the selections we applied to the \citeauthor{gil07} sample are effectively random, our final sample should be representative of the local UV- and mid-IR-bright ETG population as a whole.

\subsection{Data}\label{extphot}
We cross-matched our sample with three catalogues; the Sloan Digital Sky Survey \citep[DR7 SDSS;][\textit{u, g, r, i, {\rm and} z} filters]{yor00} the Two Micron All-Sky Survey \citep[2MASS, \K-band;][]{skr06} and WISE \citep[3.4, 4.6, 12, and 22 \micron~ filters;][]{wri10}. This provides photometry in up to 14 bands, over 0.15 to 22 \micron. The WISE point spread functions for co-added images in 3.4, 4.6, 12 and 22 \micron~ are 8.3, 9.1, 9.5, and 16.8 arcseconds, respectively. The sample comprises 25 S0 and 24 E galaxies with a mean 3.4 \micron~ radius of $R_{\rm tot} = 55$ kpc.

To obtain integrated magnitudes, we proceed as follows. For 2MASS, we obtain data from the extended source catalog \citep[2MASS XSC;][]{jar03}. For  WISE we obtain magnitudes from the extended source catalog project (WISE XS), which is described in \citet{jar13} and briefly here. For each galaxy, the total flux is integrated over an elliptical surface, determined by the convergence of the surface brightness profile and the background. The background was estimated by an outer annulus, the same minor/major-axis ratio and position angle (b/a and PA) as the elliptical aperture. Each image was visually inspected to remove foreground stars. 

For GALEX and SDSS, we follow the methods in \citet{nei11}. In the same method as \citet{nei11}, we obtain archival GALEX images to coadd for the deepest possible image. We used both GALEX imaging filters \citep{mar05}: FUV ($\rm \lambda_{eff} = 1539~\AA$) and NUV ($\rm \lambda_{eff} = 2316~\AA$).  For the SDSS data, \citet{nei09,nei11} found the SDSS catalog data to be inaccurate due to the galaxies extending over multiple strips and consequently having been broken into subregions. The authors developed a coadd and mosaic technique to obtain the proper integrated fluxes over multiple frames. We use these fluxes, and further correct for Galactic extinction, using the dust maps of \citet{schl98}.

Table \ref{tableA} includes some of the extended photometry from this analysis: ellipticity ($r$-band), WISE 3.4 \micron~ total and half-light radii. The galaxies span a wide range of total radii from $7<R_{\rm tot}<235$ kpc with half-light radii from 0.8-25 kpc. The smallest half-light radius (16 \arcsec) is bigger than the point spread function for the WISE images. The largest galaxies have total radii exceeding 100 kpc: UGC 10261 ($R_{\rm tot}/R_{\rm h} =  235/25$ kpc), NGC 7432 ($R_{\rm tot}/R_{\rm h} = 134/9.8$ kpc), and NGC 4187 ($R_{\rm tot}/R_{\rm h} = 117/12.1$ kpc). Approximately 30\% of the sample are highly elliptical (a/b $< 0.6$). The sample has a mean radial velocity, cz, of 5396 km/s. 

Five of the ETGs in our sample were observed as part of the \texttt{SAURON} 2D kinematic survey: NGC 2768, NGC 3377, NGC 4458, M 87, and NGC 5813. They have been classified as slow rotators, indicative of a quiescent stellar population \citep{mcder06}. \citet{bur11} find that the quiescent, slow rotators have blue FUV-NUV colors ($0.5\lesssim FUV-NUV \lesssim 2$). We note that the \texttt{SAURON} survey analysis is limited to a narrow spatial range within the half-light radius.

\section{Analysis\label{ssptemp}}
In this present work, we consider three possible source for the observed UV excess in ETGs: 1) a significant population of BHB/EHB stars ($\rm t_{age} >2$ Gyr and \Teff $\rm >14000$ K), where EHBs are the combined BHB phase and higher \Teff; 2) a substantial population of hot young stars from a recent small-scale starburst ($\rm t_{age} \lesssim 500$ Myr and $1<Z<1.5$ \zsun, \Teff $ > 10000$ K); 3) a significant population of very hot, post main-sequence metal-poor stars ($\rm t_{age} \gtrsim 9$ Gyr and Z $< 0.5$ \zsun). We seek to distinguish between these possibilities by comparing their predicted UV/optical/mid-IR colors. 

We use the Flexible Stellar Population Synthesis models \citep[FSPS; ][]{con09,con10} to create composite stellar population (CSP) models. The main advantage of the \citet{con09} models is the inclusion of a BHB phase, and the flexible treatments of the post-AGB and horizontal branch evolutionary phases. This flexibility allows for choosing the age at which the BHB phase is initiated, and the percentage of thermally pulsating asymptotic giant branch (TP-AGB) stars that enter that phase. In essence, adding the BHB/EHB fractions to the CSPs drives the colors blueward after \tage=2 Gyr. We combine the FSPS models with the Padova isochrones \citep{mar07,mar08} and the BaSEL stellar libraries \citep{lej97,lej98}, which provide mass, age and metallicity ranges of $0.15 \leq M \leq 100\, M_{\odot}$, $10^{6.6}<t<10^{10.2}$ yrs, and $0.005 < \mathrm{Z/Z_\odot} < 1.5$ (Z$_{\odot} \equiv 0.019$), respectively. 

For this analysis, we consider the BHB/EHB phase, and young star sources of UVX. We also include a comparison between subsolar, solar and supersolar metallicities. We limit the metallicities to 0.25, 1 and 1.5 Z$_{\odot}$, ages between 30 Myr to 14.1 Gyr at $\Delta \log(t/\mathrm{yr}) = 0.025$ intervals, apply a \citet{cha00} dust parameter of 0.3, and use the \citet{cha03} IMF.

We assume that stars on the HB are allowed to become BHB/EHB at ages $>2$ Gyr, based on \citet{yi98}. They show a UV slope increase for 1-1.5 \zsun~ populations during the post-AGB epoch, beginning at ages of $\gtrsim 2.5$ Gyr. We take three different fractions of stars on the HB that are in the BHB (or EHB) phase: 0\%, 25\% and 50\%. Note that, for all populations, we have the same total number of stars on the HB; the fraction refers to the division between `ordinary' HB stars and blue (B) or extended (E) HB stars.

The division of stars in the B/E phase between BHBs and EHBs is determined by temperature, with higher temperatures meaning more E (extended) than B (blue). The increased temperature is assumed to be due to small stellar envelope masses ($\rm <0.05 M_{\odot}$), where the radiation is allowed to escape freely. This analysis integrates temperature increases of log \delt($\log T_{\rm eff}$) from 0.0, 0.2, and 0.4 dex, which is a measure of the change from \Teff~ \citep[e.g.,][]{mar08,con09}. The parameter \delt~ is the shift in \Teff~ for a TP-AGB phase, and is included in the FSPS models based on \citet{mar08}. We chose those values based on the ranges presented for a super solar metallicity isochrone in Fig. 1 of \citet{mar08} to compare extreme boundaries in our parameter range. Finally, we apply three different e-folding times to vary the star-formation history and stellar mass with age ($\tau=$ 0.2, 0.6 and 1 Gyr). 

\fspsanalysis

We use the FSPS models to generate tracks for color analysis. Specifically, we wish to determine which combination of UV/optical/mid-IR color separates populations in as many of the following variables as possible: age (\tage), metallicity (Z), change in temperature (\delt), star formation history ($\tau$), and BHB/EHB fraction. The plots in Fig. \ref{fig:fspsanalysis} provide a sample of the explored color-space/parameter combinations.  These are an example of three color-space combinations with Z $=1$ \zsun. The top, middle and bottom rows are \nuvr/\gr,FUV-NUV/NUV-[3.4], and FUV-NUV/\nuvr, respectively. The panels in each row display the same templates, but the lines are color-coded differently to emphasize how the parameters are evolving with BHB fraction (left), e-folding time $\tau$ in Gyr (middle), and the increase in effective temperature $\Delta_T$ (log(\Teff)) at HB onset (right). For example, the left panels in each row show when the BHB fraction takes effect, when the lines  have a slower rise in \nuvr (top), or shift sharply down (blueward) in FUV-NUV (middle and bottom); the middle columns highlight the track separations for a given $\tau$ and therefore trace the star formation history (SFH); and the right panel does the same for temperature change. A more detailed description on the effects of different stellar population parameters on the UV, optical and mid-IR colors is included in Appendix \ref{sec:appendix}. 

In principle, one can do this with a UV, or UV-optical slope in the y-axis, and an optical-near-IR, or optical-mid-IR slope in the x-axis. The ultimate goal is to see which color-spaces work best; particularly extending the color-spaces into the WISE channels (from the near-IR or red optical) to determine what combination provides improved discriminatory power. We investigated several color-spaces, and determined that a UV plus WISE combination does significantly better than the UV plus optical (see Appendix \ref{sec:appendix}); we address this choice briefly in \S \ref{sec:clrclr}. 

\section{Results}\label{results}

\clrhists
\colorgrad

\subsection{Radial Color Gradients}
To study radial differences between UV, optical and mid-IR, we show color gradients for \nuvr~ and NUV-[3.4] in Fig. \ref{fig:colorgrad}. We also list, in Table \ref{tableB}, the NUV-[3.4] and \nuvr~ colors at 0.3 and 0.7 $R/R_{90}$ fractional radii for each ETG ($R_{90}$ is the radius at 90\% of the total flux in the $r$-band). There is, for almost all galaxies in our sample, a trend toward bluer colors with increasing radius. To investigate the trend, we plot in Fig. \ref{fig:clrhists} the histograms of \nuvr~ and NUV-[3.4] colors at the inner (\Rin, R within 50\% flux, white) and outer (\Rout, R between 50-90\% flux) radii. The flux-fraction is based on the total $r$-band flux for each galaxy. Both histograms show a statistically significant color separation between the peaks of approximately 1 magnitude. The averages are: \nuvr = 5.7$\pm0.2$ and 4.9$\pm0.1$ for \Rin~ and \Rout, respectively; and NUV-[3.4] = 6.1$\pm0.2$ and 5.1$\pm0.2$ for \Rin~ and \Rout, respectively. The bluest galaxy, Mrk 501, is an outlier in the sample and a known blazar. 

We also highlight subpopulations in the histograms based on ellipticity and morphological type (Fig. \ref{fig:clrhists}). Morphological type does not appear to be a significant factor in the different colors in the inner and outer regions, since their colors distribute similarly. Highly elliptical systems (b/a $<0.6$) have mean \Rin~ colors that are preferentially reddened by $\sim 2\%$ for \nuvr~ and by $\sim 3\%$ for NUV-[3.4], which is insignificant for our results. We address how dust may affect the UV to mid-IR colors in the following section.

\subsection{Comparisons to FSPS Templates}\label{sec:clrclr}

To determine the origins of the color differences, we compare the sample UV, optical and mid-IR colors of the \Rin~ and \Rout~ regions to the CSP models derived from the FSPS templates. In \S \ref{ssptemp}, we described Fig. \ref{fig:fspsanalysis} to show how changes in the parameters can dramatically alter the paths of the FSPS tracks in a given color plot. From this figure, and Appendix \ref{sec:appendix}, it can be seen that the  FUV, NUV, $r$ and [3.4] plots provide a broader range in color-space than the other optical color combinations, with which to discriminate the dominating parameters for a given color region.  For example, \gr~ and \gi~ show a color range of approximately a magnitude (Fig. \ref{fig:fspsanalysis}, and Figs. \ref{fig:clrclrA}-\ref{fig:clrclrB}), while \nuvr~ and NUV-[3.4] extend over 6 magnitudes (Fig. \ref{fig:fspsanalysis}, and Figs. \ref{fig:clrclrC}-\ref{fig:clrclrD}). Therefore, we focus on these colors in the following analysis.

\clrage

In Fig. \ref{fig:clrageA}, we show the FSPS tracks for \nuvr, NUV-[3.4], and FUV-NUV versus age ($0.2 < t_{age} < 14.1$ Gyr). The left, middle and right panels for each color-age plot are different temperature boosts given to the TP-AGB phase:  \delt~$ = 0$, 0.2, and 0.4 as labeled in the top row. The most significant effects on color are metallicity (different colors), and SFH ($\tau$). As is commonly known, higher metallicities redden the color, and a smaller $\tau$ results in a more rapid color increase. A stellar population with Z $>$ \zsun~and short e-folding time $\sim 0.2$ can become very red by \tage$=2$ Gyr.

The ETG sample colors are shown in grey and orange shaded regions that span the \Rin~ and \Rout~ values, respectively. The darker horizontal grey and orange lines mark the mean \nuvr, NUV-[3.4] and FUV-NUV values for the corresponding color shaded area. As stated previously, the inner radii have a significantly redder color than the outer radii in \nuvr~ and NUV-[3.4]. However, this reverses for FUV-NUV, where the inner color is, on average bluer by a small margin ($\sim0.1$ magnitude).  As the stellar populations age, the possible physical drivers behind the observed colors without assuming priors, quickly become degenerate. As a generalized interpretation of the color differences in Fig. \ref{fig:clrageA}, the following results are  based on the mean colors (grey and orange horizontal lines) for \Rin~ and \Rout~ at ages 2, 5 and 10 Gyr (vertical white lines).

We summarize the parameter combinations consistent with the \Rin~  and \Rout~ mean \nuvr, NUV-[3.4] and FUV-NUV colors in Table \ref{TableC}. \Rin~ tends toward higher metallicities than \Rout~ for \tage $\lesssim5$ Gyr. For both regions, $\tau$ values are limited to short e-folding times (~0.2 Gyr) at \tage $\sim 2$ Gyr; the e-folding time is degenerate at 5 and 10 Gyr. The BHB fraction is impossible to isolate at any temperature or age in these plots. However, the inner mean color is consistent with having little to no BHB population.

Additionally, it appears that temperature is only important at the extreme end \delt~$=0.4$. Despite the degeneracies, we find indications that age is primarily responsible for the color difference in our ETG sample. Just from the three sampled ages, we show the progression of the different modeled stellar populations that might lead to color differences. 

To  explore the premise that age is the main cause of the color differences in the ETGs, we take isochrones of the synthetic photometry in Fig. \ref{fig:clrageA} at \tage = 2, 5, and 10 (white vertical lines in Fig. \ref{fig:clrageA}, and plot them in FUV-NUV, \nuvr, NUV-[3.4] color-space. We do this in Fig. \ref{fig:isochrones}  by plotting FSPS lines on top of the grey-scale contours that include both the \Rin~and \Rout~ colors of the ETG sample. We outline the distribution of colors for \Rout~ with the dashed contour lines. We also add dust vectors (orange arrows crossing the Z$= 1$ and 1.5 \zsun~ tracks) with an increasing dust parameter from 0-1 using the \citet{cha00} dust models. The top, middle and bottom rows are the isochrones at ages 2, 5 and 10 Gyr as labeled.

Our first assessment is that the density peaks are clearly separated, and dust may be ruled out simply by observing the direction of the vectors in Fig. \ref{fig:isochrones}. If dust was preferentially reddening the inner half-light of the ETGs, the FUV-NUV \Rin~ colors would lie redward. We do not mean that these are entirely dustless (we actually include dust in the FSPS models; see \S \ref{ssptemp}). Rather, we assume that dust is not a strong enough factor to cause the color difference. 

The 2 Gyr plots (top row in Fig. \ref{fig:isochrones}, indicate that the colors for the inner regions could only be explained by super-solar metallicities $>1.5$ \zsun. At the same age, the outer regions are more likely to host a 1-1.5 \zsun~ stellar population with a moderate BHB fraction. At 5 Gyr (middle plots), the \Rin~ colors could be caused by higher metallicity/moderate BHB fractions, and the outer regions would have a lower metallicity and higher numbers of  BHBs. At 10 Gyr,  the isochrones give narrower ranges and the metallicity dependence is stark: \Rin~ would be dominated by 1-1.5 \zsun~ stellar population and an insignificant BHB fraction; \Rout~ would be dominated by a $<1$ \zsun~ stellar population, and may include slightly more BHBs. We also note that for 2-5 Gyr the dominating SFH is $\tau = $ 0.2-0.6; at 10 Gyr the range is widened to all $\tau$. 

The above conditions indicate two possible scenarios: 1) if the bulk of the bulge and disk mass coevolved, the metallicities must be different by $\sim1$ \zsun; 2) the ETGs formed in an inside-out process with at least 2 major stages of growth $\gtrsim 1$ Gyr apart. These appear to be two distinctly different mechanisms of  formation, and we deepen our analysis in the following section to determine the likelihood of either scenario.

\subsection{Age estimates for \Rin~ and \Rout}\label{sec:ageclr}

To constrain the likelihood of the results based on Fig. \ref{fig:isochrones}, we estimate the ages for the inner and outer regions of the ETGs. However, we showed in the previous section that the high degeneracy of the synthetic photometry makes it difficult to derive accurate ages for each of the objects (e.g., Fig. \ref{fig:clrageA}). In order to check the results, given the degeneracies, we resample the data to find the most likely statistical fit to the ages given a fixed set of parameters. We fit the data to the FSPS templates, with the FUV-NUV, \nuvr, and NUV-[3.4] colors. We resampled the inner and outer colors by replacement estimates (bootstrapping) within the errors for 500 iterations to estimate errors and probabilities for the $\chi^2$ fitting. The $\chi^2$ fitting was done between the resampled observed colors and the FSPS data. After minimizing the $\chi^2$ for each iteration, we use the maximum likelihood for each parameter combination. The probability distribution functions were used to estimate the weighted average age. The ages and errors were calculated for each BHB, $\tau$, Z and \delt~ combination for each object.

\isochrones

In Fig. \ref{fig:chisqage} we show the effects that BHB fraction, metallicity and star formation history have on the estimated ages (\tage(in) and \tage(out)). Each data point is a weighted average of the entire ETG sample, testing different parameter configurations (e.g., one point for \fbhb$= 0 $, $ \tau = 0.2$, and Z$=1$ \zsun). We do not separate the temperature boosts here, since Fig. \ref{fig:clrageA} indicates that the effects on color are negligible for this analysis.

The points reflect the assumption that both the inner and outer regions are evolving with the same set of parameters. We are essentially testing one condition for the first scenario in the previous section \S \ref{sec:clrclr}. For example, the magenta square (Z=1.5\zsun, \fbhb=0.5, and $\tau=1$), easily identified as the outlier, is fixing the prior that the parameter combinations are the same for the inner and outer regions. In other words, each point is the \tage coordinate for the inner and outer regions having the same metallicity, BHB fraction, and $\tau$. With this assumption, we find that nearly all points are below the equality line, and many (12 points) with $\Delta$ \tage $>1$ Gyr. 

In order for the inner and outer regions to have equal ages, then the inner and outer regions must have a certain set of properties. Over half of the points within errors of the equal age line, have low metallicity (double circles for Z=0.25 \zsun) and some fraction of BHBs (diamonds and squares). This constrains our hypothetical scenarios in the previous section.
  
 Average ages derived from low metallicity (0.25 \zsun) show the least difference between \tage(in)  and \tage(out). The average age separation is 0.28, 1.13, and 0.9 Gyr for Z=0.25, 1, and 1.5 \zsun, respectively. When selected by BHB fraction, the average differences are 1.21, 0.86, and 0.24 Gyr for \fbhb = 0, 0.25, and 0.5, respectively; $\tau$ shows little difference 0.83, 0.8, and 0.69 Gyr for $\tau=$ 0.2, 0.6 and 1, respectively. 
 
Given these priors, 67\% of the ages for the inner regions are older (to the right of the shaded region), and have an average age difference of 1.1 Gyr. The mean difference for all points is $0.8$ Gyr with a range of -0.3 to 1.9 Gyr if the parameters are assumed to be equal. The average age for the inner region is 7.0 $\pm 0.3$ Gyr and the outer is 6.2 $\pm 0.2$ Gyr. The assumptions we used to derive these results indicate that the outer regions are likely to have formed 0.5-1.2 Gyr after the inner regions. 
\chisqage
\ageZratios

\subsection{Age-Metallicity Ratios}\label{sec:ageZgradients}
We explore the possibility of an age-metallicity difference between \Rin~ and \Rout~ in our sample. Since we are limited to 3 metallicities, the results we present are a crude estimate of how metallicity and age behave between these regions. Fig. \ref{fig:ageZratios} shows the inner to outer average age ratios for different $\rm Z_{inner} \, and \, Z_{outer}$ metallicity combinations. The error bars span the range of ages for the set of parameters at that metallicity ratio (e.g., BHB, $\tau$). We fit the $\rm \Delta \log (Age)$-$\rm \Delta \log (Z)$ values with the following result:

 \begin{eqnarray} 
   \rm \Delta \log (Age) = -0.16 \Delta \log (Z) \nonumber.
 \end{eqnarray} 

Fig. \ref{fig:ageZratios} shows that, for $\rm Z_{inner}/Z_{outer} < 1$, it is more likely that the stellar populations in the inner half-lights of our sample are older than the outer even though the range of ages spans widely about the mean. For $\rm Z_{inner}/Z_{outer} > 1$, the ages span a narrow range around $\rm t_{inner}/t_{outer} =1$.  For metallicity gradients of decreasing Z with radius, the estimated age difference for the ETGs increases, resulting in an average of 1.4 Gyr (older toward the centers), which is nearly double the average when we assumed $\rm Z_{inner}/Z_{outer} = 1$ (see \S \ref{sec:ageclr}). This further supports the hypothesis that the color difference are likely due to multi-stage evolution and less of a metallicity difference. However, the information presented here is not conclusive of a presence, or lack thereof, of a metallicity gradient.

\subsection{BHBs and the color NUV-[3.4] color difference}\label{sec:clrclrbhb}
The majority of our analysis is dedicated to determining whether the color difference is indicative of co-evolving or multi-stage evolution. We also want to know the source of UV emission in these ETGs, since we have shown our sample is unlikely to have had recent star formation. The age estimates indicate that the significant UV emission is largely produced by sources in an older stellar population (\tage$>2$ Gyr; see Figs. \ref{fig:isochrones} and \ref{fig:chisqage}). One question we asked is whether a significant fraction of BHB stars can accomplish this. Can the overall trend of bluer NUV-[3.4] and \nuvr~ with increasing galactocentric radius be due to the presence of $\gtrsim 25\%$ BHBs in a 2-5 Gyr-old stellar population? For example, in Fig. \ref{fig:isochrones}, at 10 Gyr the BHB fraction is important only within a small area of color-space: $0.6<$FUV-NUV$<1.5$, $4.4<$\nuvr $<5.7$, $4.2<$NUV-[3.4] $<6.3$. The mean \Rin/\Rout~ values are: (FUV-NUV) = 1.2/1.3; (\nuvr) = 5.7/4.9; and (NUV-[3.4]) = 6.1/5.1, which do fall into these ranges of color-space, indicating that BHBs could be a source of the UV emission.

\bhbclrdiffs

To investigate how they may contribute to the 1 magnitude color difference between \Rin~ and \Rout, we assign different BHB fractions between the inner and outer radii and extract the corresponding NUV and [3.4] photometry from FSPS templates, using the estimated ages for the ETG sample at \Rin~ and \Rout~ (see \S \ref{sec:ageZgradients}). Fig. \ref{fig:bhbclrdiffs} shows the stacked distributions of $\rm \Delta (NUV-[3.4]) = (NUV-[3.4])_{inner} - (NUV-[3.4])_{outer})$. Higher \fbhb~ in the outer regions are most likely to make colors redder in the centers. In this case, a jump of \fbhb $=0.25$ in the outer region produces this effect. 

\subsection{Mass, Star Formation and UV-mid-IR Color }
We explore if the colors in the inner and outer radii are related to star formation rates and stellar mass (SFR and $\rm M^*$, respectively) derived from integrated photometry (e.g., for low surface brightness, or high redshift sources). We first compute star formation rates and stellar masses from the integrated magnitudes (for the whole system in each case), using the UV calibration of \citet[Eq. 1]{ken98a} and the 2MASS \K~ vs. \gr~ calibration of \citet{bel03}. 

\massclr

In Fig. \ref{fig:massclr} we plot these values against the NUV-[3.4] colors for the individual sources, and find: 1) the SFRs show a scatter for both the inner and outer colors; 2) there is an increase in stellar mass with redder color. This increase in mass with more red ETGs is consistent with other observations \citep[e.g.,][]{bel03}. The binned averages are the total flux colors (not inner and outer, but the entire galaxy), averaged over SFR and stellar mass bins. The SFR is very low for all of the galaxies; therefore, we can assume the bluer colors for the outer radii are not created by recent star formation. The majority of the ETGs are within the massive galaxy range ($>10^{11} \rm M_\odot$). We note that the \citet{bel03} analysis does not include contributions of HB stars to the UV. However, it is useful to discern if the color difference could be caused by a primarily young stellar population.

\section{Discussion}\label{discussion}
We have shown that there exists a clear difference in the UV to IR color between the inner and outer regions of our sample of 49 E/S0 galaxies. Moreover, we see no significant variation between the E and S0 Hubble types. In this section, we discuss which stellar mass assembly histories are consistent with these observations based on our analysis in the previous section. The CSP predictions of color based on BHBs, ages and metallicities, illustrate that secular coevolution of a bulge and disk progenitor would be difficult to explain without a significant internal mixing mechanism. 

The first result we found was the 1 magnitude difference between the colors measured for the inner half-light and and the outer radii. We hypothesized that a different stellar population must be dominating the respective regions, since color gradients in ETGs and globular clusters have been repeatedly measured. For example, \citet{car11} use GALEX and 2MASS to discuss the radial UVX in ETGs; they find a steep gradient between the inner and outer regions with an increase in UVX at the core. They attribute this to increased $\alpha$ enhancement from Type II SNe, and perhaps He abundance, rather than to dry mergers in an outside-in scenario. Moreover, recent papers by \citet{for11} and \citet{arn11} present optical color radial profiles for a nearby globular cluster NGC 1407 and S0 galaxy NGC 3115, respectively. Both show evidence of a two-stage formation history, using color analysis with \gi. They find a color difference and decrease in metallicity [Fe/H] between the inner and outer regions. 

We took advantage of the better-defined separations of the FSPS tracks in the NUV-[3.4] color-space, and augmented this with the \nuvr~ and FUV-NUV color, to narrow down the possibilities between the well-known degeneracies. In the density plots (Fig. \ref{fig:isochrones}),  the separate peaks for inner and outer fall in regions depending on the assumed age, and support that EHBs are the possible source of UV emission. Furthermore, this is consistent with being driven primarily by differences in age, with secondary contributions from differences in BHB fraction, metallicity, and star formation history.

Based on these assessments, and as we mentioned in \S \ref{sec:clrclr}, two possible formation scenarios are consistent with comparison of the FSPS colors and the different colors of the inner and outer regions in the ETG sample: 1) if the bulge and disk coevolved, the metallicities must be different by $\gtrsim 1$ \zsun; 2) the ETGs formed in an inside-out process with at least 2 major stages of growth $\gtrsim 1$ Gyr apart.

For the first case, the inner parts of the galaxy would need to exhibit a strong metallicity gradient within a short time-frame through self-enrichment or merging.  If the hierarchical model accurately describes the formation of the bulk of the bulge mass, then the timescales for relaxation are on order of 0.1-1 Gyr \citep[e.g., ][]{bar96}, and pseudobulge formation at timescales $>2$ Gyr. Surveys of globular clusters in M31 show a mix of metal-poor to metal-rich clusters across the galaxy with metal-rich clusters predominately in the central 10 kpc radius \citep{sai00,perrett02}. \citet{sai00} model the timescale to produce a suitable metallicity gradient over the galactocentric radius, and calculate that self-enrichment with a collapsing disk would take $\sim5$ Gyr, while a merging event would take $\sim2$ Gyr. Both surveys conclude that merging with low-metallicity galaxies is most likely to have produced the metallicity gradients, supporting multi-stage evolution. If this is typical for most early type galaxies, evidence of a second-stage or multi-stage event would be expected. The timescales of self-enrichment are too long for the inner and outer stellar populations to have coevolved, which leads us to the second scenario.

Our results are more consistent with an inside-out multi-stage evolution. Whether or not the color difference between the inner and outer radii is caused by a metallicity gradient, the centers of the ETGs must be on average $\sim1$ Gyr older, according to our results. The question remains what merger process, gas rich or poor, is most likely to produce the age/color differences observed.

In principle, several processes can contribute to stellar mass assembly in local massive galaxies. First, \textit{in situ } star formation may be triggered by accretion of gas from the IGM. Mergers may also play a part through `wet' (dissipational) mergers between gas-rich galaxies, or `dry' (dissipationless) mergers between gas-poor galaxies. The stellar ages for wet mergers tend to be younger, supplying the gas for starbursts. A `frosting' of gas-rich mergers is often used to explain the dispersion in ages within ETGs \citep{tra00b,san06,san09}. For dry mergers, the most massive bulge-dominated ETGs form by the accretion of galaxies with older stellar populations onto the central bulge. 

Each of these may occur via major or minor mergers, and observational evidence varies on which is most influential in terms of mass building \citep[e.g.,][]{vandok05,cox06,san09}. The analysis in \citet{vandok05} shows that 71\% of bulge-dominated galaxies have some signature of interaction (i.e., tidal disturbances), and 35\% have had a recent dry major merger. \citet{san09} did a spectral analysis of the \citet{vandok05} sample and find that the sample easily divides into a range of wet to dry merger stages in age and metallicity. The metallicities are 1-1.5 \zsun~ for all merger scenarios \citep{san06,san09}, but the dry mergers tend toward slightly higher metallicities and ages $\gtrsim 5$ Gyr. Simulations by \citet{cox06} show that wet mergers leave remnants that are smaller, with higher rotation, and a more disk-like shape. Boxy, massive ellipticals can result from dry mergers, but the observed nearby ETGs are likely formed by a combination of wet/dry merging. Our sample consists of more massive ellipticals, and the estimated ages ($\sim7$ Gyr) support dry merging for $z\lesssim0.3$. If a metallicity gradient exists, dry-merging with higher metallicity galaxies may be consistent with our age-Z trends. We find inconsistent trends with our first scenario, where we proposed that higher metallicities at the center would cause the red colors. However, our calculations will need to be verified with a more rigorous age-Z radial profile fitting.

There is evidence that the contribution from wet and dry mergers may result in a distinct two-stage formation history for ETGs. Observations show that the inner and outer regions of nearby galaxies decouple at some point, implying that a wet process is producing transitions and velocity dispersions indicative of recent star formation \citep{treu05,ems07,kor09,arn11,for11}. In an inside-out evolution, the oldest stellar populations exist in the center and the outer halo or disk has or had more recent star formation via minor gas rich mergers.

Recent observations from \citet{nel12} support inside-out evolution for rapidly forming disks at $z\sim1$, and agree with our age and multi-stage formation time-frames. Their results show an extended H$\alpha$ disk, outside of an older central bulge, compared with the R-band stellar continuum ($r_{e}{\rm (H\alpha)} > 1.3 \,r_{e}({\rm R})$). \citet{dam09} compare the restframe optical radii of compact passive galaxies at $1 <z<2$ with nearby ETGs. They show that the inner cores of the nearby ETGs are consistent with the sizes of the higher redshift compact galaxies. 

Our results are compatible with a two- or multi-stage, inside-out formation history, in which the core formed through rapid star formation, via wet mergers or cold gas accretion from the IGM at least $\sim 7$ Gyr ago ($z\gtrsim $2). After  $\sim 1-2$ Gyr (at $z\sim 1$), a second phase of minor-merging would induce a new major starburst phase possibly in an outer disk of the galaxy. Our age estimates require that the inner and outer regions are both greater than 2 Gyr old, and differ by an average of 1.4 Gyr. 

Additionally, the outer regions are consistent with having an elevated fraction of EHBs that can begin as early as 2-5 Gyr, based on the synthetic photometry (see Fig. \ref{fig:clrageA}). Our results show that an increase of 0.25 BHB fraction in the outer regions can produce the 1 magnitude color difference we observe. The BHB/EHB phases are predicted to have lifetimes of $\sim 10$ Myr, so a population of stars in this phase would be very short-lived compared to the average age of $\gtrsim 4$ Gyr in the inner regions.

A possible alternative explanation is differential extinction between inner and outer regions. However the extinction vectors (orange arrows) shown in Fig. \ref{fig:isochrones}  imply they would need to be over a magnitude of extinction, and the FUV-NUV would show the opposite separation between the inner and outer regions, which we do not observe. It is possible but unlikely, though we cannot completely rule it out from our data. This may explain why we see a slight bias for the highly elliptical galaxies (b/a$<0.6$) at the inner radii (see Fig. \ref{fig:clrhists}).

\section{Conclusions}

The longstanding question of what causes UVX is likely not answered by one simple population (old or young) or parameter tweak. We explore the origin of the UV emission in our sample using stellar population synthesis models. We also chose within a range of values, the parameters (BHB fraction, Z, $\tau$, \delt) that best fit the UV, optical and mid-IR colors of our sample. We then looked at the spatial breakdown of UV to optical and UV to mid-IR colors. Through this approach, the following results are observed:

\begin{itemize}
\item WISE and GALEX colors FUV-NUV and NUV-[3.4] are highly effective in separating the parameters which drive the observed colors. Our CSP tracks show that, at 10 Gyr, the BHB fraction is important only within a small area of color-space: $0.6<$FUV-NUV$<1.5$, $4.4<$\nuvr $<5.7$, $4.2<$NUV-[3.4] $<6.3$.

\item  The 49 ETGs in this sample exhibit a strong color difference with bluer colors on the outside of the galaxies in both UV-opt and UV-mid-IR. We extract the photometry for the inner half-light and outer 50-90\% and find a clear color difference independent of E/S0 type. The average value of the ETG \Rin/\Rout~ colors are \nuvr = 5.7/4.9, and NUV-[3.4] = 6.1/5.1.

\item Increasing the fraction of BHBs in the outer regions by 0.25 can create the observed color difference, where the inner are redder than the outer regions by 1 magnitude in NUV-[3.4].

\item We find that the properties of the inner and outer regions are significantly different, either in age, metallicity, and/or the existence of BHBs. We discuss two formation scenarios based on our results: 1) if the bulge and disk coevolved, the metallicities must be significantly different ($0.25 $ \zsun in the outer regions, $>1$ \zsun in the centers); 2) the ETGs formed in an inside-out process with at least 2 major stages of growth $\gtrsim 1$ Gyr apart. Our age estimates indicate that the second scenario is most likely. 

\item  The average ages are estimated to be 7.0 $\pm 0.3$ Gyr (inner) and 6.2 $\pm 0.2$ Gyr (outer), with a minimum of 2.6 Gyr. Even when we assume homogeneity of parameters over radius, there is evidence of multi-stage evolution where the outer regions are likely to have formed at least $\sim0.8$  Gyr after the inner regions (-0.3 $<\langle\Delta  $\tage$\rangle <$ 1.9 Gyr). 

\item The age differences estimated for metallicity gradients with an increase in Z at larger radial distances are younger at the center by an average of $\sim 300 $ Myr. For increasing Z with radius they present age differences (\tage (in) - \tage (out)) of 1.4 Gyr with a range of 0.6-2.2 Gyr. This suggests that age and other stellar population properties are contributing to the color difference rather than metallicity.

\item Since the estimated ages are beyond the lifetimes of star forming regions to contribute significantly to the UV emission, we assume that BHBs or EHBs are the primary source of the UVX, agreeing with previous results \citep[e.g., ][]{gre90}. The average colors fall within the ranges predicted for BHB fractions greater than 0.25 in NUV-[3.4] in the outer regions.

\end{itemize}

These combined results lead us to the conclusion that the UV observed in these massive ETGs is likely caused by an extreme HB phase of older stars; and that the ETGs went through a multi-stage evolution that is coincident with an inside-out cessation. We do not discount that star formation is causing UV in other ETGs, but our particular selection may have biased us toward more massive ETGs that are slow-rotators and not star-forming. A larger sample will be explored in detail in a subsequent paper to look at the individual properties of the galaxies and their environments in more detail. Previous studies do not probe the range of ETGs type with the radially binned extended photometry, and these very extended objects won't have the coverage out to large radii WISE naturally provides.

\acknowledgements 
We thank the anonymous referee for their thorough comments, which greatly improved this paper. We thank Marcio Catelan for discussions on evolved stellar populations. Thank you to D. Stern for numerous insights to the discussion on galaxy evolution. We also thank R. Assef for his feedback on the analysis and early development of this project. This publication makes use of data products from the Wide-field Infrared Survey Explorer, which is a joint project of the University of California, Los Angeles, and the Jet Propulsion Laboratory/California Institute of Technology, funded by the National Aeronautics and Space Administration. The publication is based on observations made with the NASA Galaxy Evolution Explorer. GALEX is operated for NASA by the California Institute of Technology under NASA contract NAS5-98034. This publication makes use of data products from the Sloan Digital Sky Survey. Funding for the SDSS and SDSS-II has been provided by the Alfred P. Sloan Foundation, the Participating Institutions, the National Science Foundation, the U.S. Department of Energy, the National Aeronautics and Space Administration, the Japanese Monbukagakusho, the Max Planck Society, and the Higher Education Funding Council for England. The SDSS Web Site is http://www.sdss.org/. This publication makes use of data products from the Two Micron All Sky Survey, which is a joint project of the University of Massachusetts and the Infrared Processing and Analysis Center/California Institute of Technology, funded by the National Aeronautics and Space Administration and the National Science Foundation. This research has made use of the NASA/IPAC Extragalactic Database (NED) which is operated by the Jet Propulsion Laboratory, California Institute of Technology, under contract with the National Aeronautics and Space Administration.

\tableA
\tableB
\tableC

%%%%%%%%%%%%%Appendix FSPS Analysis
\appendix

{\twocolumn
\section{Color Analysis of UVX in Stellar Populations}\label{sec:appendix}

We used the FSPS photometry to determine the best approach for comparing UV-optical, optical-optical, optical-mid-IR and UV-mid-IR colors.The analysis was carried out as described in \S \ref{ssptemp}. We briefly recap the CSP model parameters here. We combine the FSPS models with the Padova isochrones \citep{mar07,mar08}, and limit the metallicities to 0.25, 1 and 1.5 Z$_{\odot}$, ages between 30 Myr to 14.1 Gyr at $\Delta \log(t/\mathrm{yr}) = 0.025$ intervals, apply a \citet{cha00} dust parameter of 0.3, and use the \citet{cha03} IMF. 

For BHBs, we assume that stars on the HB are allowed to become BHB/EHB at ages $>2$ Gyr, taking three different fractions of stars on the HB that are in the BHB (or EHB) phase: 0\%, 25\% and 50\%. To address the EHB phase, assuming these are BHBs with higher effective temperatures, we include temperature boosts \delt~= 0.2, 0.4 dex.

In Figs. \ref{fig:clrclrA}, \ref{fig:clrclrB}, \ref{fig:clrclrC}, and \ref{fig:clrclrD}, we show how colors evolve with changes in the parameters, using \nuvr/\gr, \nuvg/\gi, FUV-NUV/\nuvr, FUV-NUV/NUV-[3.4]. Each panel from left to right shows the same FSPS tracks, but the lines are colored differently based on parameter values to emphasize the dependence of that parameter on the position in color-space (see labels in the top row). For example, the top row in Fig. \ref{fig:clrclrA} shows: (left) the BHB fractions of 0, 0.25 and 0.5 depicted in black, blue and brown, respectively; (middle) the star formation history $\tau =$ 0.2, 0.6 and 1 Gyr in black, blue and brown, respectively; (right) the change in effective temperature \delt~ at 0, 0.2, and 0.4 dex in black, blue and brown, respectively. The consecutive rows show how the FSPS tracks change with Z as indicated in each panel.
\clrclrA
\clrclrB
\clrclrC
\clrclrD
Separation in color-space for each variable is not strong for \gr, \nuvg, and \gi. All three have smaller ranges, and smaller separations, leading to significant degeneracies. Concerning the former point, the color ranges in the panels in Fig. \ref{fig:clrclrA} are very narrow, spanning 0.7 in optical, compared with the \nuvr~ and \nuvg~ ($\sim 6$). For the second point, the largest difference between the BHB and $\tau$ lines for \gi~ and \gr~ are $\lesssim 0.1$, and 0.3, respectively. Separation in $\tau$ is reasonable, but separation in BHB fraction and temperature is relatively poor except at very old ages. 

We extend our analysis to show the same tracks (with the same color-coding) in two examples of color-spaces that combine UV and mid-IR data; FUV-NUV/NUV-[3.4] and FUV-NUV/\nuvr~ (Fig. \ref{fig:clrclrB}). These show a much wider span in color on the x-axis by 5 magnitudes, compared to the 0.7 magnitude range of the \nuvr/\gr~ and \nuvg/\gi~ plots (Fig. \ref{fig:clrclrA}). 

Additionally, there is a much cleaner separation in color for $\tau$, and somewhat cleaner separations for both BHB and temperature. The GALEX-WISE colors provide an expanded range in color-space for BHB fraction color-cuts, especially at high metallicity, and there is a marginal improvement for changes in \delt. For example, comparing the top left plots in Fig. \ref{fig:clrclrA} and \ref{fig:clrclrB}, the different BHB fractions are separated on average by one magnitude in the GALEX-WISE, but only 0.2 magnitudes in the GALEX-SDSS. A further advantage of NUV-[3.4] colors is the near all-sky coverage of both datasets. Because of these advantages, we adopted the FUV-NUV/NUV-[3.4] and FUV-NUV/\nuvr~ spaces as the primary color-space focus throughout the paper. 
\clearpage

}
\clearpage

\ajbib

\end{document}

%% file: ajbib.tex
\def\ajbib{
}

%% file: figs-tables_spetty_etgs_v7-pp2.tex
\def\tableA{
\begin{deluxetable}{lllcccccc}
\tablewidth{0pt}
\tabletypesize{\scriptsize}
\tablecaption{Early type galaxy sample properties \label{tableA}}

\tablehead{\colhead{Object} & 
\colhead{RA} & 
\colhead{DEC} & 
\colhead{b/a\tablenotemark{a}} & 
\colhead{$R_{\rm tot}$\tablenotemark{b}} & 
\colhead{$R_{\rm h}$\tablenotemark{b}} & 
\colhead{c$z$\tablenotemark{c}} & 
\colhead{Type\tablenotemark{e}} \\ 
\colhead{} & 
\colhead{} & 
\colhead{} & 
\colhead{} & \colhead{(kpc)} & 
\colhead{(kpc)} & 
\colhead{(km/s)} & 
\colhead{} }

\startdata
NGC 155  & 00:34:40.0  & -10:45:58.7   &   0.87 & 75   & 8.1  &  6210	&   S0 \\
NGC 163  & 00:35:59.7  & -10:07:17.5    &   0.93 & 60   & 7.7  &  5982	&   E  \\
IC 1698  & 01:25:22.0  & 14:50:19.7    &   0.41 & 27   & 8.7  &  6503	&   S0 \\
IC 1700  & 01:25:24.5  & 14:51:52.8    &   0.81 & 58   & 8.6  &  6356	&   E  \\
UGC 1040 & 01:27:35.8  & -01:06:18.3     &   0.25 & 32   & 6.3  &  4489	&   S0-a  \\
NGC 1047 & 02:40:32.7  & -08:08:51.4     &   0.55 & 13   & 1.8  &  1340	&   S0-a  \\
NGC 1052 & 02:41:04.8  & -08:15:21.0    &   0.71 & 32   & 1.9  &  1510	&   E4 \\
NGC 1060 & 02:43:15.0  & 32:25:30.0    &   0.74 & 86   & 7.2  &  5190	&   S0 \\
NGC 1066 & 02:43:49.9  & 32:28:29.7    &   0.71 & 62   & 5.8  &  4346	&   E  \\
NGC 1361 & 03:34:17.7  & -06:15:54.0    &   0.68 & 50   & 6.0  &  5272	&   E  \\
UGC 4188 & 08:03:24.0   & 41:54:53.3    &   0.50 & 108  & 13.9 &  9685	&   S0 \\
UGC 4551 & 08:44:06.0  & 49:47:37.6    &   0.31 & 13   & 2.4  &  1749	&   S0 \\
NGC 2675 & 08:52:04.9  & 53:37:02.3    &   0.69 & 103  & 12.3 &  9231	&   E  \\
IC 522   & 08:54:34.9  & 57:10:00.7     &   0.81 & 42	& 7.1  &  5079   &   S0 \\
NGC 2693 & 08:56:59.2  & 51:20:50.7     &   0.69 & 76	& 6.6  &  4942   &   E3 \\
UGC 4702 & 08:58:51.2  & 38:48:34.4    &   0.93 & 76   & 11.1 &  8443	&   S0 \\
NGC 2768 & 09:11:37.4  & 60:02:14.2      &  0.40  & 31	&1.7   &  1373   &   E \\
NGC 3265 & 10:31:06.7  & 28:47:47.7    &   0.73 & 8	& 1.7  &  1319   &   S0 \\
NGC 3377 & 10:47:42.3  & 13:59:09.0     &  0.47  & 16	&1.0   &  665	 &   E5 	\\
UGC 5928 & 10:49:47.1  & 51:53:39.0    &   0.98 & 59   & 9.8  &  7393	&   S0 \\
NGC 3522 & 11:06:40.4   & 20:05:07.6     &   0.53 & 12   & 1.1  &  1221	&   E  \\
NGC 3539 & 11:09:08.8   & 28:40:19.9     &   0.33 & 74	& 12.8 &  9707   &   S0-a  \\
UGC 6435 & 11:25:35.0  & 00:46:05.5    &   0.78 & 99   & 10.1 &  7604	&   S0 \\
UGC 6683 & 11:43:16.2  & 19:44:55.4    &   0.35 & 39   & 9.2  &  7542	&   S0-a  \\
NGC 3844 & 11:44:00.8  & 20:01:45.6      &  0.26  & 46	&8.9   &  6779   &   S0-a  \\
NGC 4187 & 12:13:29.3  & 50:44:29.0    &   0.63 & 117  & 12.1 &  9138	&   E  \\
NGC 4458 & 12:28:57.6  & 13:14:30.9    &   0.93 & 7	& 0.8  &  635	 &   E0 \\
NGC 4478 & 12:30:17.4  & 12:19:42.4    &   0.81 & 15   & 1.8  &  1349	&   E2 \\
M 87 	 & 12:30:49.4  & 12:23:28.0    &   0.93 & 52   & 1.8  &  1307	&   E  \\
IC 3457  & 12:31:51.4  & 12:39:25.0    &   0.68 & 17   & 2.1  &  1297	&   E3 \\
NGC 4787 & 12:54:05.5  & 27:04:07.0     &   0.32 & 52   & 10.4 &  7585	&   S0-a  \\
NGC 4797 & 12:54:55.1  & 27:24:45.7      &  0.74  & 84   &10.1  &  7863   &   S0	\\
NGC 4827 & 12:56:43.5  & 27:10:43.2    &   0.79 & 61   & 9.9  &  7630	&   E  \\
NGC 4952 & 13:04:58.3   & 29:07:19.7     &   0.60 & 58   & 8.2  &  5968	&   E  \\
NGC 5004 & 13:11:01.5  & 29:38:12.1     &   0.72 & 57	& 9.3  &  7044   &   S0 \\
NGC 5173 & 13:28:25.3  & 46:35:29.5    &   0.96 & 22   & 3.2  &  2419	&   E0 \\
NGC 5329 & 13:52:10.0  & 02:19:30.4    &   1.00 & 102  & 9.7  &  7109	&   E  \\
UGC 8986 & 14:04:15.8   & 04:06:44.7     &   0.98 & 9	& 1.8  &  1232   &   S0 \\
NGC 5576 & 14:21:03.6  & 03:16:15.4    &   0.66 & 30   & 2.0  &  1487	&   E3 \\
NGC 5638 & 14:29:40.3  & 03:14:00.5    &   0.87 & 20   & 2.3  &  1676	&   E  \\
IC 1024  & 14:31:27.1  & 03:00:32.2      &   0.43 & 15	& 2.0  &  1454   &   S0 \\
IC 1071  & 14:54:12.4  & 04:45:00.9    &   0.63 & 107  & 10.2 &  8302	&   S0 \\
NGC 5813 & 15:01:11.2   & 01:42:07.2    &   0.66 & 51   & 2.6  &  1972	&   E  \\
UGC 10261 & 16:11:04.0 & 52:27:02.1    &   0.68 & 235  & 25.0 &  19018  &   S0 \\
Mrk 501  & 16:53:52.2  & 39:45:36.6    &   0.78 & 107  & 13.3 &  10092  &   E  \\
NGC 6364 & 17:24:27.3  & 29:23:25.4     &  0.78  & 74	&9.3   &  6874   &   S0 	\\
UGC 10935 & 17:38:43.2 & 57:14:21.6    &   0.51 & 49   & 12.7 &  8801	&   S0 \\
NGC 7317  & 22:35:51.8 & 33:56:41.8    &   0.95 & 56   & 8.2  &  6599	&   E4 \\
NGC 7432  & 22:58:01.9 & 13:08:04.4      &  0.59  & 134  &9.8   &  7615   &   E \\

\enddata
\tablenotetext{a}{Ellipticities at the total major and minor axes are based on the extended photometry measurements in the SDSS $r$-band.}
\tablenotetext{b}{Total and half-light radii come from the WISE 3.4 \micron~ photometry.}
\tablenotetext{c}{Radial velocities were obtained from the NASA Extragalactic Database archive (http://ned.ipac.caltech.edu/).}
\tablenotetext{d}{The Hubble type, according to the \cite{gil07} UV catalog.}
\end{deluxetable}
}

\def\tableB{
\begin{deluxetable}{lcccc}
\tablewidth{0pt}
\tabletypesize{\scriptsize}
\tablecaption{UV-optical and UV-mid-IR colors at specific radii \label{tableB}}

\tablehead{\colhead{Object} & 
\colhead{NUV-[3.4]\tablenotemark{a}} & 
\colhead{NUV-[3.4]\tablenotemark{b} } & 
\colhead{NUV-$r$\tablenotemark{a} } & 
\colhead{NUV-$r$\tablenotemark{b}} \\ 
\colhead{} & 
\colhead{$R/R_{90}=0.3$} &
\colhead{$R/R_{90}=$0.7} & 
\colhead{$R/R_{90}=$0.3} & 
\colhead{$R/R_{90}=$0.7} } 
\startdata
NGC 155  & 6.23$\pm$0.03 & 5.81$\pm$0.02 & 6.09$\pm$0.03 & 5.58$\pm$0.02 \\
NGC 163  & 6.37$\pm$0.02 & 6.15$\pm$0.02 & 6.05$\pm$0.02 & 5.77$\pm$0.01 \\
IC 1698 & 6.00$\pm$0.04 & 5.14$\pm$0.02 & 5.61$\pm$0.04 & 4.64$\pm$0.02 \\
IC 1700 & 6.53$\pm$0.03 & 6.07$\pm$0.02 & 6.17$\pm$0.03 & 5.72$\pm$0.02 \\
UGC 1040 & 6.85$\pm$0.05 & 6.52$\pm$0.03 & 6.76$\pm$0.05 & 6.28$\pm$0.03 \\
NGC 1047 & 4.33$\pm$0.01 & 4.55$\pm$0.01 & 4.38$\pm$0.01 & 4.53$\pm$0.01 \\
NGC 1052 & 6.42$\pm$0.01 & 6.10$\pm$0.01 & 5.89$\pm$0.01 & 5.67$\pm$0.01 \\
NGC 1060 & 7.79$\pm$0.03 & 7.31$\pm$0.03 & 6.88$\pm$0.03 & 6.43$\pm$0.03 \\
NGC 1066 & 7.84$\pm$0.06 & 7.13$\pm$0.05 & 6.91$\pm$0.06 & 6.25$\pm$0.04 \\
NGC 1361 & 6.16$\pm$0.03 & 5.80$\pm$0.03 & 5.93$\pm$0.03 & 5.49$\pm$0.03 \\
UGC 4188 & 7.05$\pm$0.05 & 6.30$\pm$0.03 & 6.75$\pm$0.05 & 5.78$\pm$0.03 \\
UGC 4551 & 6.64$\pm$0.02 & 6.41$\pm$0.02 & 6.53$\pm$0.02 & 6.04$\pm$0.01 \\
NGC 2675 & 6.63$\pm$0.04 & 6.37$\pm$0.03 & 6.64$\pm$0.04 & 5.98$\pm$0.02 \\
IC 522 & 6.85$\pm$0.04 & 5.93$\pm$0.02 & 6.58$\pm$0.04 & 5.52$\pm$0.02 \\
NGC 2693 & 6.12$\pm$0.01 & 6.05$\pm$0.01 & 5.75$\pm$0.01 & 5.64$\pm$0.01 \\
UGC 4702 & 6.55$\pm$0.04 & 6.05$\pm$0.03 & 6.27$\pm$0.04 & 5.65$\pm$0.03 \\
NGC 2768 & 6.74$\pm$0.01 & 6.26$\pm$0.01 & 6.18$\pm$0.01 & 5.87$\pm$0.0 \\
NGC 3265 & 3.59$\pm$0.01 & 3.66$\pm$0.01 & 3.34$\pm$0.01 & 3.39$\pm$0.01 \\
NGC 3377 & 6.18$\pm$0.01 & 5.57$\pm$0.01 & 5.75$\pm$0.01 & 5.33$\pm$0.0 \\
UGC 5928 & 6.09$\pm$0.05 & 5.78$\pm$0.03 & 6.09$\pm$0.05 & 5.58$\pm$0.03 \\
NGC 3522 & 5.60$\pm$0.02 & 5.12$\pm$0.01 & 5.90$\pm$0.02 & 5.17$\pm$0.01 \\
NGC 3539 & 7.36$\pm$0.09 & 6.49$\pm$0.04 & 7.00$\pm$0.09 & 5.96$\pm$0.04 \\
UGC 6435 & 6.46$\pm$0.03 & 5.96$\pm$0.02 & 6.31$\pm$0.03 & 5.63$\pm$0.02 \\
UGC 6683 & 6.59$\pm$0.05 & 6.42$\pm$0.04 & 6.61$\pm$0.05 & 6.17$\pm$0.03 \\
NGC 3844 & 6.91$\pm$0.05 & 6.54$\pm$0.03 & 6.72$\pm$0.05 & 6.20$\pm$0.03 \\
NGC 4187 & 6.81$\pm$0.03 & 6.35$\pm$0.02 & 6.25$\pm$0.02 & 5.93$\pm$0.02 \\
NGC 4458 & 5.88$\pm$0.01 & 5.47$\pm$0.01 & 5.77$\pm$0.01 & 5.41$\pm$0.01 \\
NGC 4478 & 6.38$\pm$0.01 & 6.07$\pm$0.01 & 6.17$\pm$0.01 & 5.85$\pm$0.01 \\
M 87 & 5.55$\pm$0.01 & 5.65$\pm$0.01 & 5.19$\pm$0.01 & 5.35$\pm$0.01 \\
IC 3457 & 4.62$\pm$0.01 & 4.56$\pm$0.01 & 4.85$\pm$0.01 & 4.71$\pm$0.01 \\
NGC 4787 & 6.56$\pm$0.05 & 5.97$\pm$0.03 & 6.53$\pm$0.05 & 5.70$\pm$0.03 \\
NGC 4797 & 6.61$\pm$0.01 & 6.25$\pm$0.01 & 6.04$\pm$0.01 & 5.79$\pm$0.01 \\
NGC 4827 & 6.20$\pm$0.01 & 6.04$\pm$0.01 & 6.05$\pm$0.01 & 5.68$\pm$0.01 \\
NGC 4952 & 6.28$\pm$0.02 & 6.11$\pm$0.02 & 6.19$\pm$0.02 & 5.75$\pm$0.01 \\
NGC 5004 & 6.37$\pm$0.02 & 6.06$\pm$0.01 & 6.24$\pm$0.02 & 5.69$\pm$0.01 \\
NGC 5173 & 4.61$\pm$0.01 & 4.31$\pm$0.01 & 4.67$\pm$0.01 & 4.23$\pm$0.01 \\
NGC 5329 & 6.29$\pm$0.01 & 5.99$\pm$0.01 & 6.01$\pm$0.01 & 5.68$\pm$0.01 \\
UGC 8986 & 5.19$\pm$0.01 & 4.81$\pm$0.01 & 5.32$\pm$0.01 & 4.96$\pm$0.01 \\
NGC 5576 & 6.05$\pm$0.01 & 5.71$\pm$0.01 & 5.86$\pm$0.01 & 5.52$\pm$0.01 \\
NGC 5638 & 6.35$\pm$0.01 & 5.96$\pm$0.01 & 5.99$\pm$0.01 & 5.71$\pm$0.01 \\
IC 1024 & 4.75$\pm$0.02 & 4.32$\pm$0.01 & 3.64$\pm$0.01 & 3.64$\pm$0.01 \\
IC 1071 & 6.98$\pm$0.02 & 6.51$\pm$0.02 & 6.37$\pm$0.02 & 5.98$\pm$0.02 \\
NGC 5813 & 6.47$\pm$0.01 & 6.14$\pm$0.01 & 6.00$\pm$0.01 & 5.72$\pm$0.01 \\
UGC 10261 & 6.63$\pm$0.02 & 6.21$\pm$0.01 & 6.09$\pm$0.01 & 5.65$\pm$0.01 \\
Mrk 501 & 2.56$\pm$0.01 & 3.09$\pm$0.01 & 2.18$\pm$0.01 & 2.46$\pm$0.01 \\
NGC 6364 & 6.71$\pm$0.02 & 6.27$\pm$0.01 & 6.47$\pm$0.02 & 5.91$\pm$0.01 \\
UGC 10935 & 6.92$\pm$0.08 & 5.89$\pm$0.04 & 6.39$\pm$0.08 & 5.43$\pm$0.04 \\
NGC 7317 & 6.96$\pm$0.02 & 6.78$\pm$0.02 & 6.74$\pm$0.02 & 6.18$\pm$0.02 \\
NGC 7432 & 6.63$\pm$0.03 & 6.29$\pm$0.02 & 6.21$\pm$0.03 & 5.82$\pm$0.02 \\
\enddata
\tablenotetext{a}{Colors for NUV-[3.4] and \nuvr~ measured at the $r$-band fractional radii $R/R_{90}=0.3$. See Fig. \ref{fig:colorgrad}.}
\tablenotetext{b}{Colors for NUV-[3.4] and \nuvr~ measured at the $r$-band fractional radii $R/R_{90}=0.7$. See Fig. \ref{fig:colorgrad}.}

\end{deluxetable}
}

 \def\tableC{\begin{deluxetable}{cccc|ccc|ccc}

\tabletypesize{\footnotesize}
\tablewidth{0pt}
\tablecaption{Possible parameter solutions for the mean \Rin~ and \Rout~ colors at 2, 5 and 10 Gyr \label{TableC}}
\tablehead{
\colhead{}& \multicolumn{3}{c|}{\delt = 0} &\multicolumn{3}{c|}{\delt = 0.2}  &\multicolumn{3}{c}{\delt = 0.4}  \\  
 \cline{2-4} \cline{5-7} \cline{8-10}
\colhead{$t_{\rm age}$} & \colhead{Z (\zsun)} & \colhead{$\tau$ (Gyr)} & \colhead{$f_{\rm BHB}$}  \vline&\colhead{Z (\zsun)} & \colhead{$\tau$ (Gyr)} & \colhead{$f_{\rm BHB}$} \vline& \colhead{Z (\zsun)} & \colhead{$\tau$ (Gyr)} & \colhead{$f_{\rm BHB}$} } 

\startdata
\hline 
\Rin &  &  &  &  &  &  &  &  &  \\
2 Gyr & 1.5 & 0.2 & $<0.25$ & 1.5 & 0.2 & $<0.25$ & \nodata & \nodata & \nodata \\
5 Gyr & 1-1.5 & 0.6 & 0-0.5 & 1-1.5 & 0.2-0.6 & $<0.25$ & 1-1.5 & 0.2-0.6 & $<0.25$ \\
10 Gyr & 0.25-1.5 & 0.2-1 & 0-0.5 & 1-1.5 & 0.2-1 & $<0.25$ & 1-1.5 & 0.2-1 & $ <0.25$ \\
\hline
\Rout &  &  &  &  &  &  &  &  &  \\
2 Gyr & 1-1.5 & 0.2 & 0-0.5 & 1-1.5 & 0.2 & $>0.25$ & 1-1.5 & 0.2 & 0-0.5 \\
5 Gyr & 0.25-1.5 & 0.2-0.6 & 0-0.5 & 0.25-1 & 0.2-0.6 & 0-0.5 & 0.25-1 & 0.2-0.6 & 0-0.5 \\
10 Gyr & $<1$ & 0.2-1 & 0-0.5 & 0.25-1 & 0.2-1 & 0-0.5 & 0.25-1 & 0.2-1 & 0-0.5 \\
\enddata

\end{deluxetable}

}
\def\fspsanalysis{
\begin{figure*}
\epsscale{1.2}
 \plotone{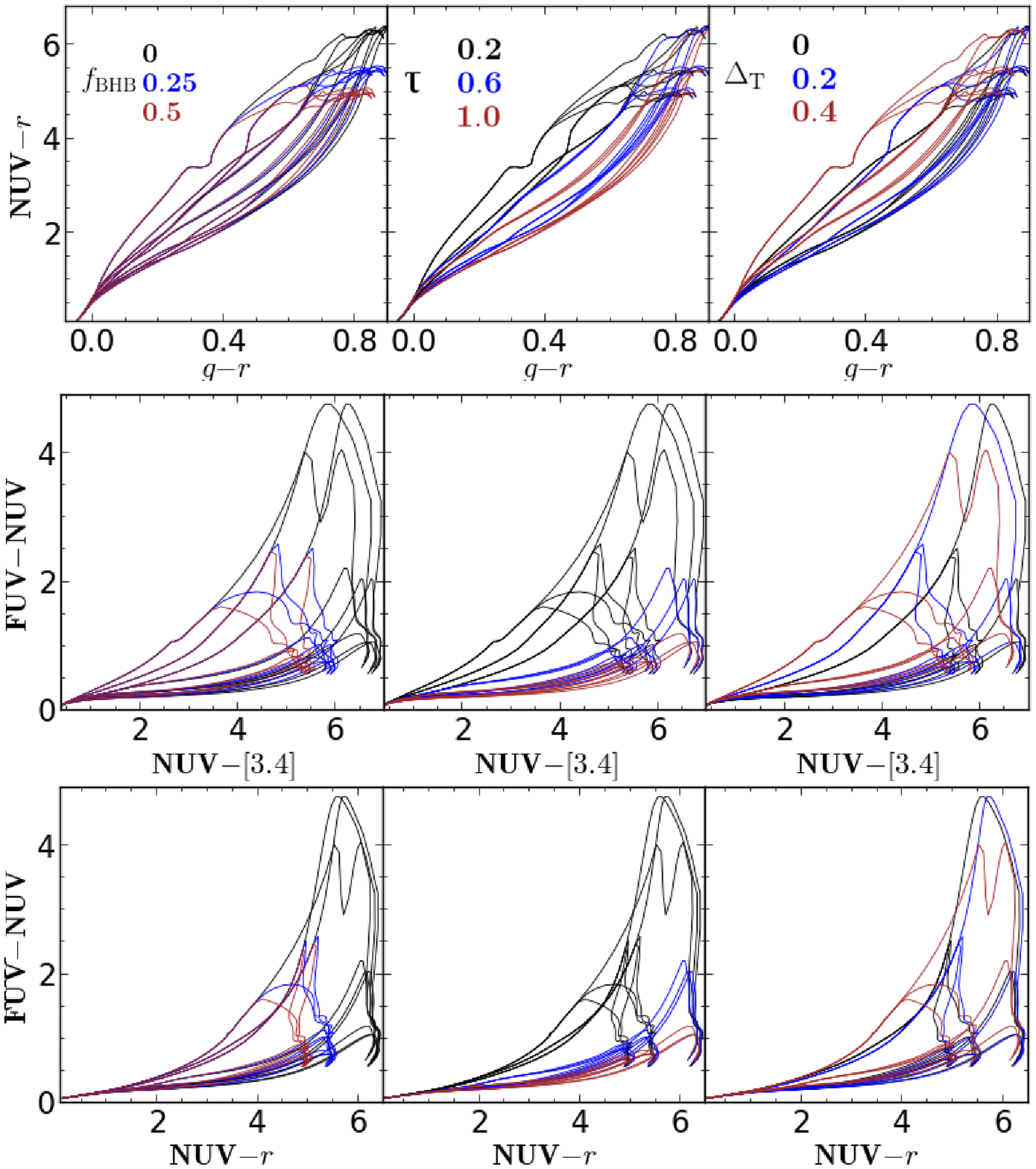}
 \caption{Flexible Stellar Population Synthesis \citep[FSPS,][]{con09}, composite stellar population (CSP) templates depicting the evolution of color-space within an age range of 0.03-14.1 Gyr. The three rows show \nuvr/\gr~ (top), FUV-NUV/NUV-[3.4] (middle), and FUV-NUV/\nuvr~ (bottom) all with Z $=1$ \zsun. The left/middle/right panels in each row show the same templates/color-color values. We color-code the lines differently (left to right, as labeled in the top row) to emphasize how the parameters are evolving with BHB fraction, e-folding time $\tau$ in Gyr, and the increase in effective temperature $\Delta_T$ (log(\Teff)) at HB onset.
  }\label{fig:fspsanalysis}
 \end{figure*}
}

\def\colorgrad{
\begin{figure*}
\epsscale{1}
 \plotone{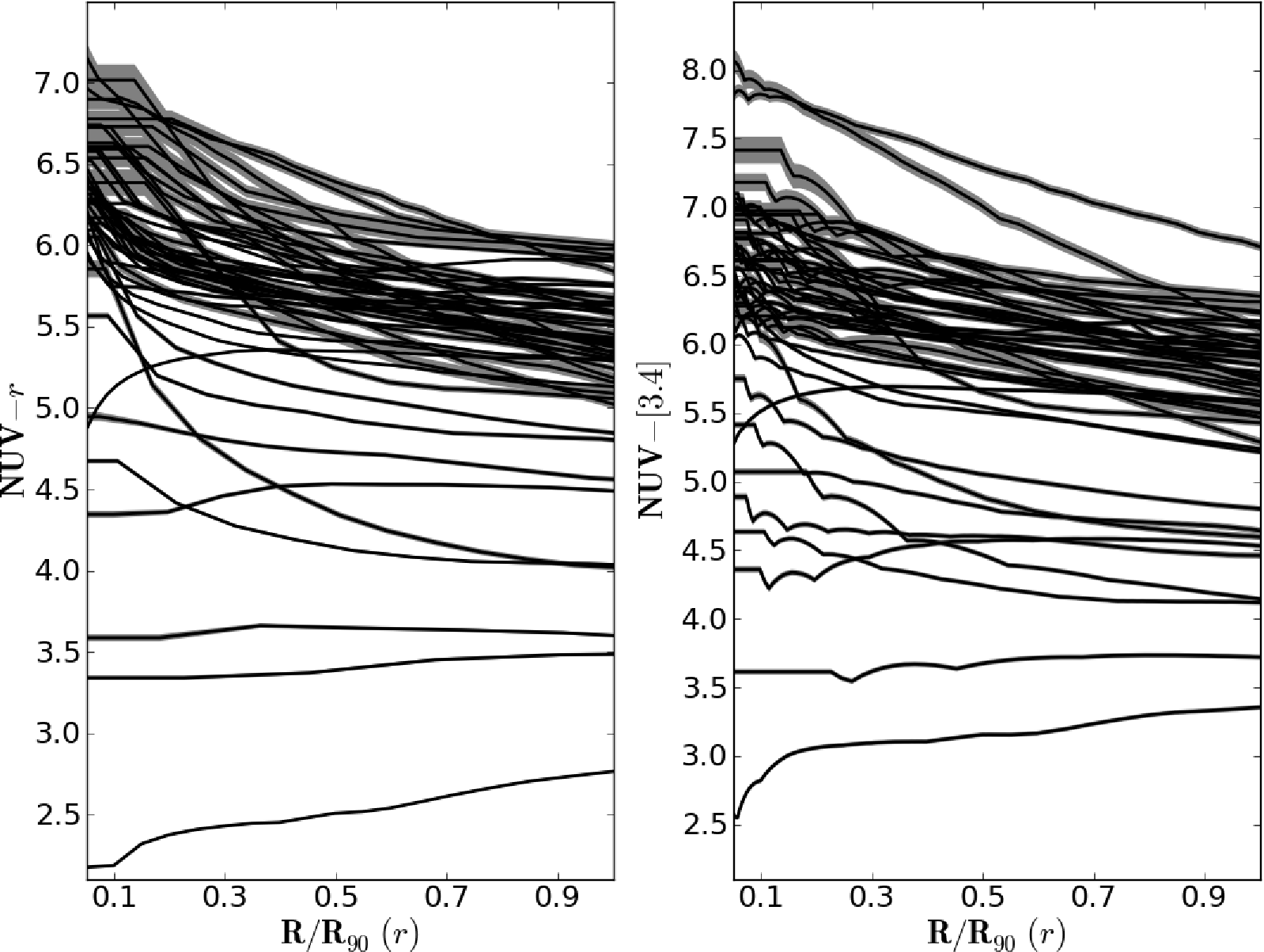}
\caption{Color as a function of fractional $r$-band radii: \nuvr~ (left) and NUV-[3.4] (right). The fractional radius is taken at $R/R_{90}$, where $R_{90}$ is the radius at 90\% of the total flux in the $r$-band. There is a distinct trend toward bluer colors outward of the half-light radius. The grey shading is the estimated error. Table \ref{tableB} lists NUV-[3.4] and \nuvr~ values at 0.3 and 0.7 $R/R_{90}$.
}
\label{fig:colorgrad}
\end{figure*}

}

\def\clrhists{
\begin{figure*}
\epsscale{1}
\plotone{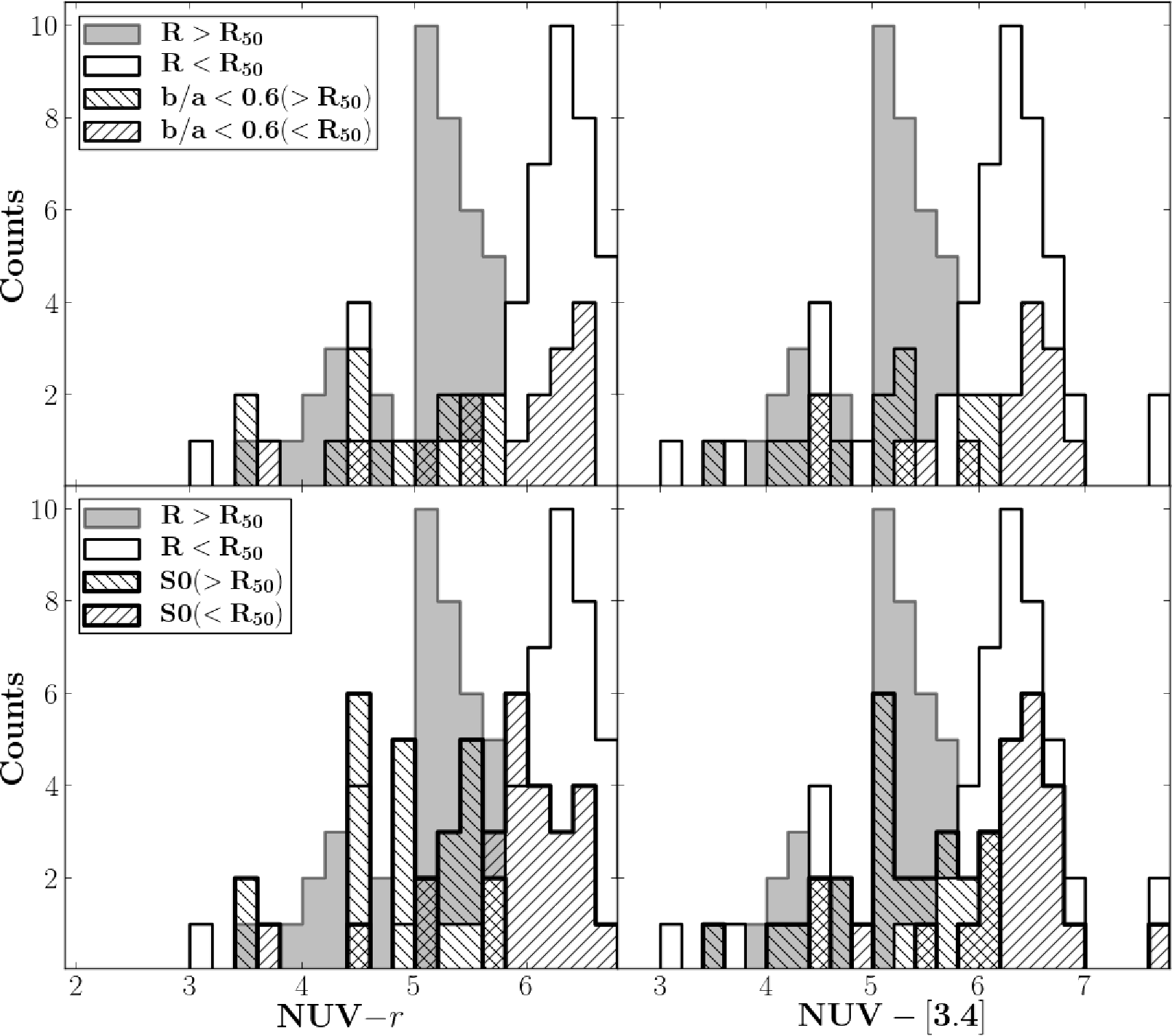}
\caption{Distributions of \nuvr~ and NUV-[3.4] colors at the inner (\Rin, R within 50\% flux, white) and outer (\Rout, R between 50-90\% flux, grey) radii. The hatches in the top panels accent a subpopulation of highly elliptical galaxies (b/a $<0.6$); the bottom  highlight S0 galaxies. There is a distinct color separation between the inner and outer colors, agreeing with the trend seen in Fig. \ref{fig:colorgrad}. This separation occurs at NUV-[3.4] $\gtrsim 6$, which is a strong color-cut for galaxies with low BHB fraction and 1-1.5 \zsun (see \S \ref{ssptemp} and Appendix \ref{sec:appendix}, Fig. \ref{fig:clrclrD}).
}
\label{fig:clrhists}
\end{figure*}
}

\def\clrage{
\begin{figure*}

\epsscale{1.8}
\plotone{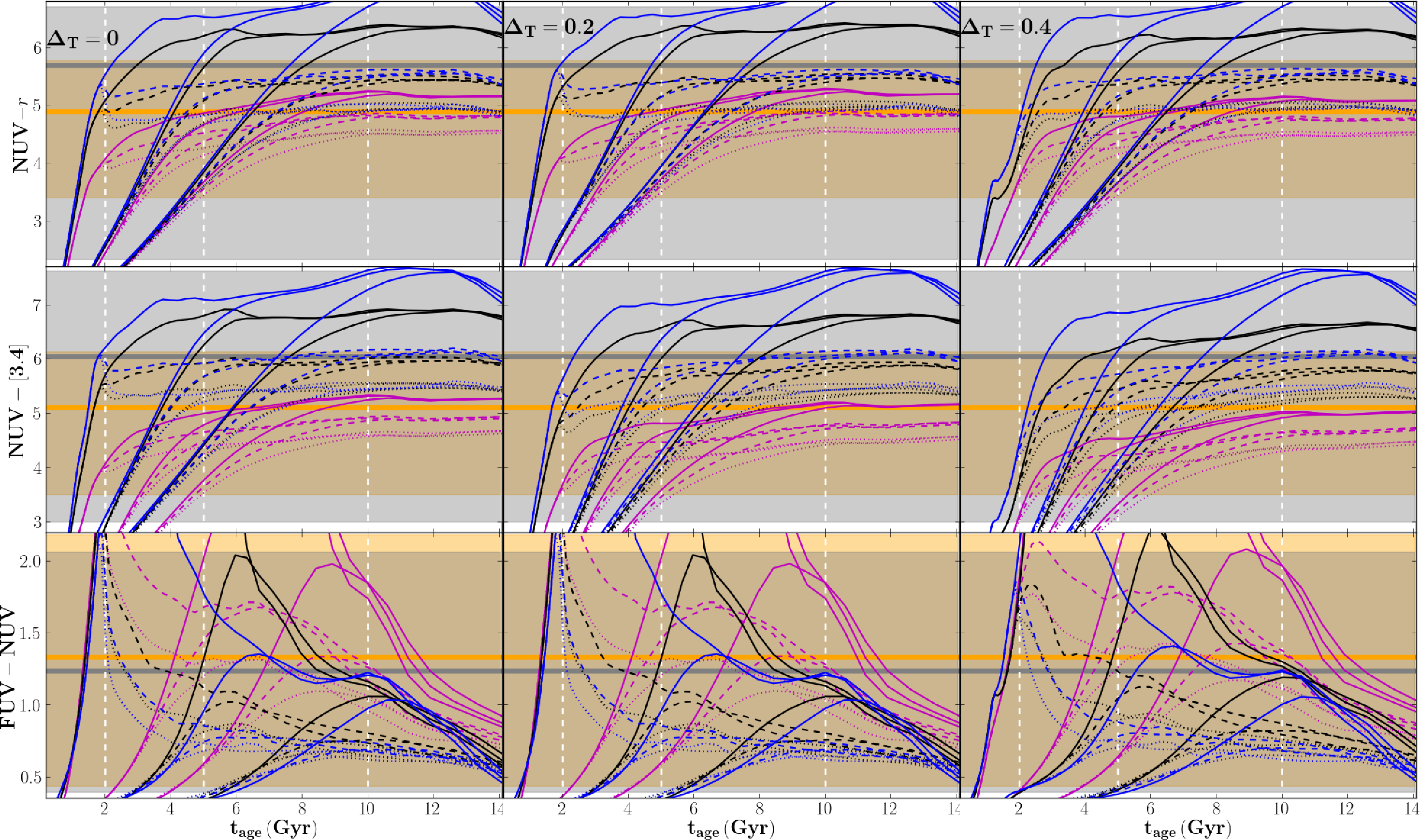}
\caption{To illustrate the significant degeneracies with age, we plot color vs. age from $0.2 < t_{age} < 14.1$ Gyr, with the ETG sample (orange and grey shading), and the FSPS colors (\nuvr, NUV-[3.4], FUV-NUV) with temperature boosts \delt $ = 0$, 0.2, and 0.4 from left to right (as labeled). The magenta, black and blue lines are Z $= 0.25$, 1.0, and 1.5 \zsun, respectively. The solid, dashed, and dotted lines represent the 0, 0.2 and 0.5 BHB fractions initiated at $t_{age} = 2$ Gyr. Each linestyle has three lines (not labeled) for the three star formation histories ($\tau = 0.2$, 0.6, 1.0). The range of color for the inner and outer regions of the ETGs are shaded in grey and orange bands, respectively. The solid horizontal lines indicate the mean values for the respective regions. The white vertical lines are placed at 2, 5 and 10 Gyr, which we inspect in Fig. \ref{fig:isochrones}.
}\label{fig:clrageA}
\end{figure*}
}

\def\clrclrA{
\begin{figure*}
\epsscale{1.5}
 \plotone{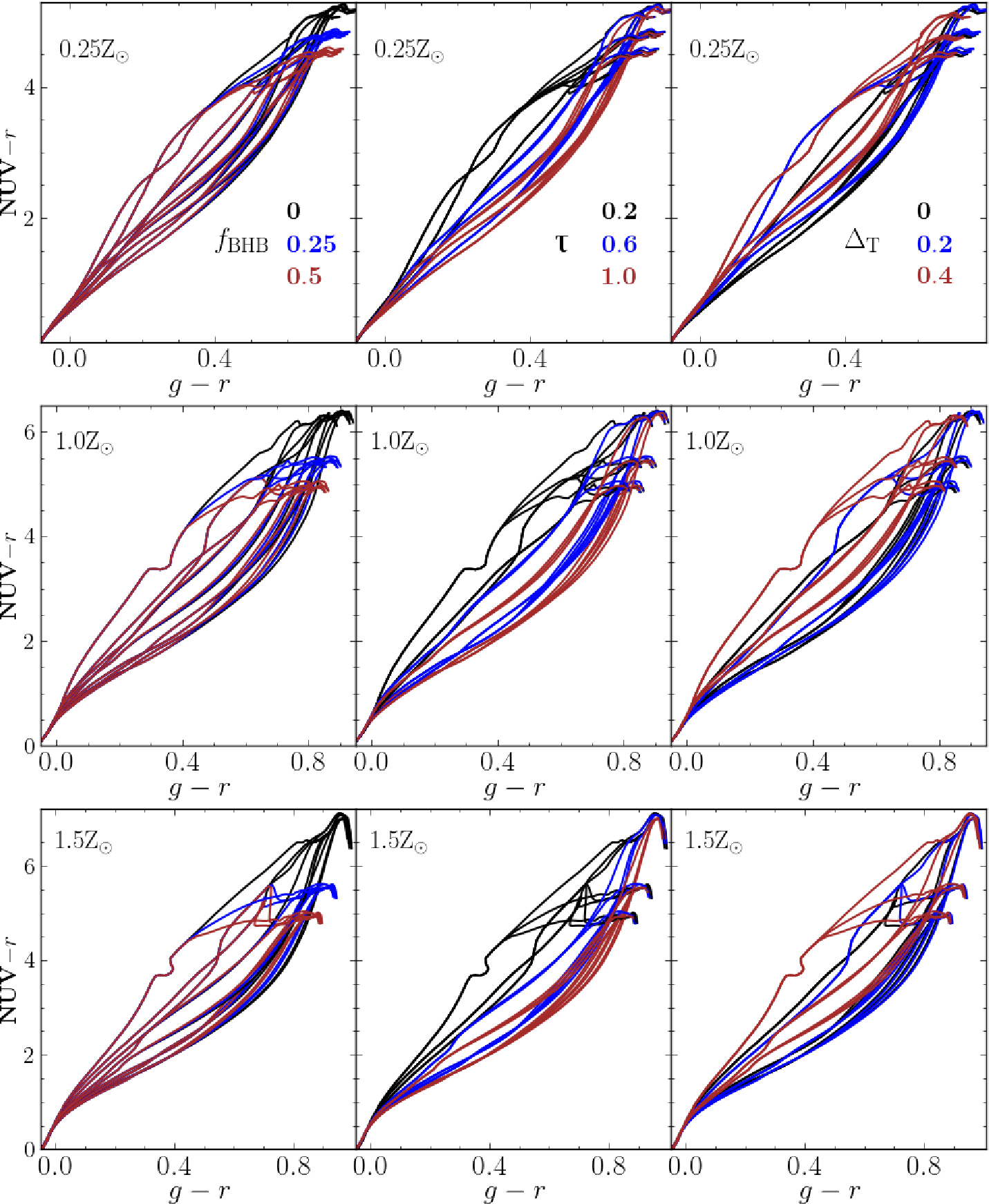}
 \caption{Flexible Stellar Population Synthesis \citep[FSPS,][]{con09}, composite stellar population (CSP) templates depicting the evolution of \nuvr~ vs. \gr~ color-space within an age range of 0.03-14.1 Gyr. The three rows depict tracks for different metallicities Z$= 0.25$, 1.0 and 1.5 \zsun as labeled. Each row/metallicity shows the same templates/color-color values. The lines are colored differently (left to right) to emphasize how the parameters are evolving as labeled in the top row panels: BHB fraction, e-folding time $\tau$ in Gyr, and the increase in effective temperature $\Delta_T$ (log(\Teff)) at HB onset. 
  }\label{fig:clrclrA}
 \end{figure*}
}

\def\clrclrB{
 \begin{figure*}
\epsscale{1.5}
 \plotone{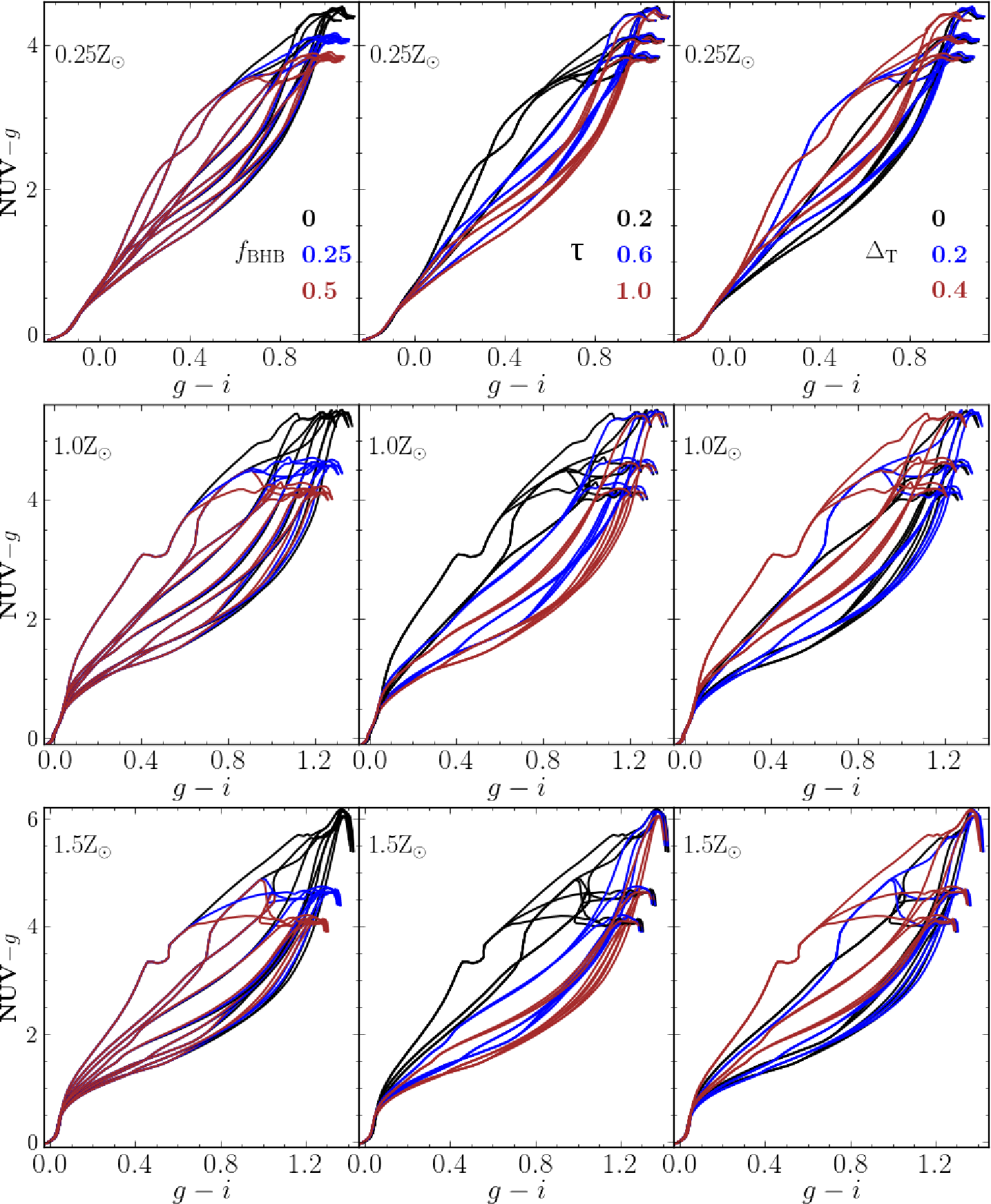}
 \caption{FSPS CSP templates depicting the evolution of \nuvg~ vs. \gi~ color-space from ages 0.03-14.1 Gyr. See Fig. \ref{fig:clrclrA} for a full description.
 }\label{fig:clrclrB}
 \end{figure*}
}

\def\clrclrC{
 \begin{figure*}
\epsscale{1.5}
 \plotone{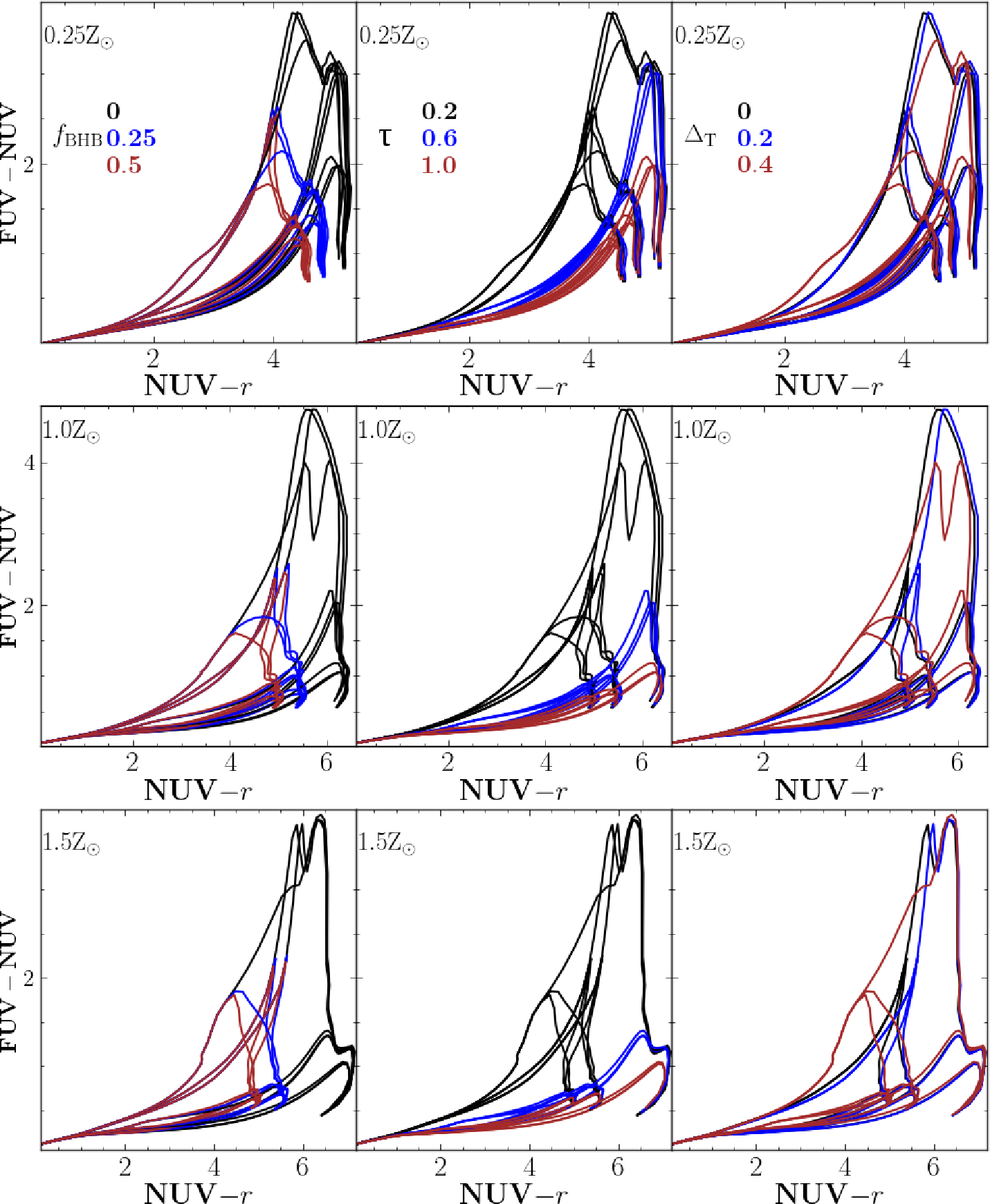}
 \caption{FSPS CSP templates depicting the evolution of FUV-NUV vs. \nuvr~ color-space from ages 0.03-14.1 Gyr. See Fig. \ref{fig:clrclrA} for a full description.
 }\label{fig:clrclrC}
 \end{figure*}
}

\def\clrclrD{
 \begin{figure*}
\epsscale{1.5}
 \plotone{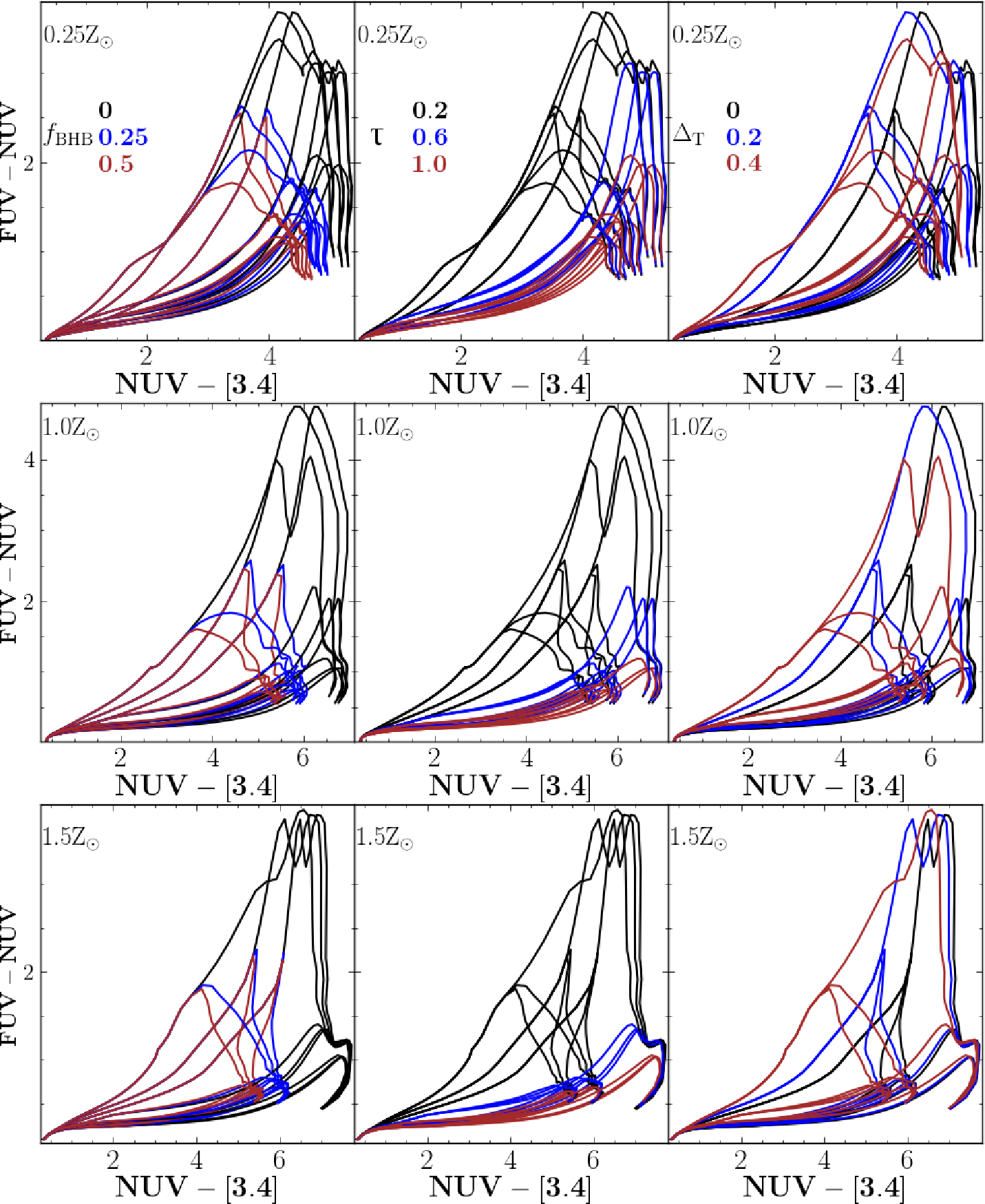}
 \caption{FSPS CSP templates depicting the evolution of FUV-NUV vs. NUV-[3.4] color-space with from ages 0.03-14.1 Gyr. See Fig. \ref{fig:clrclrA} for a full description.
 }\label{fig:clrclrD}
 \end{figure*}
}

\def\isochrones{
\begin{figure*}
\epsscale{1}
\plotone{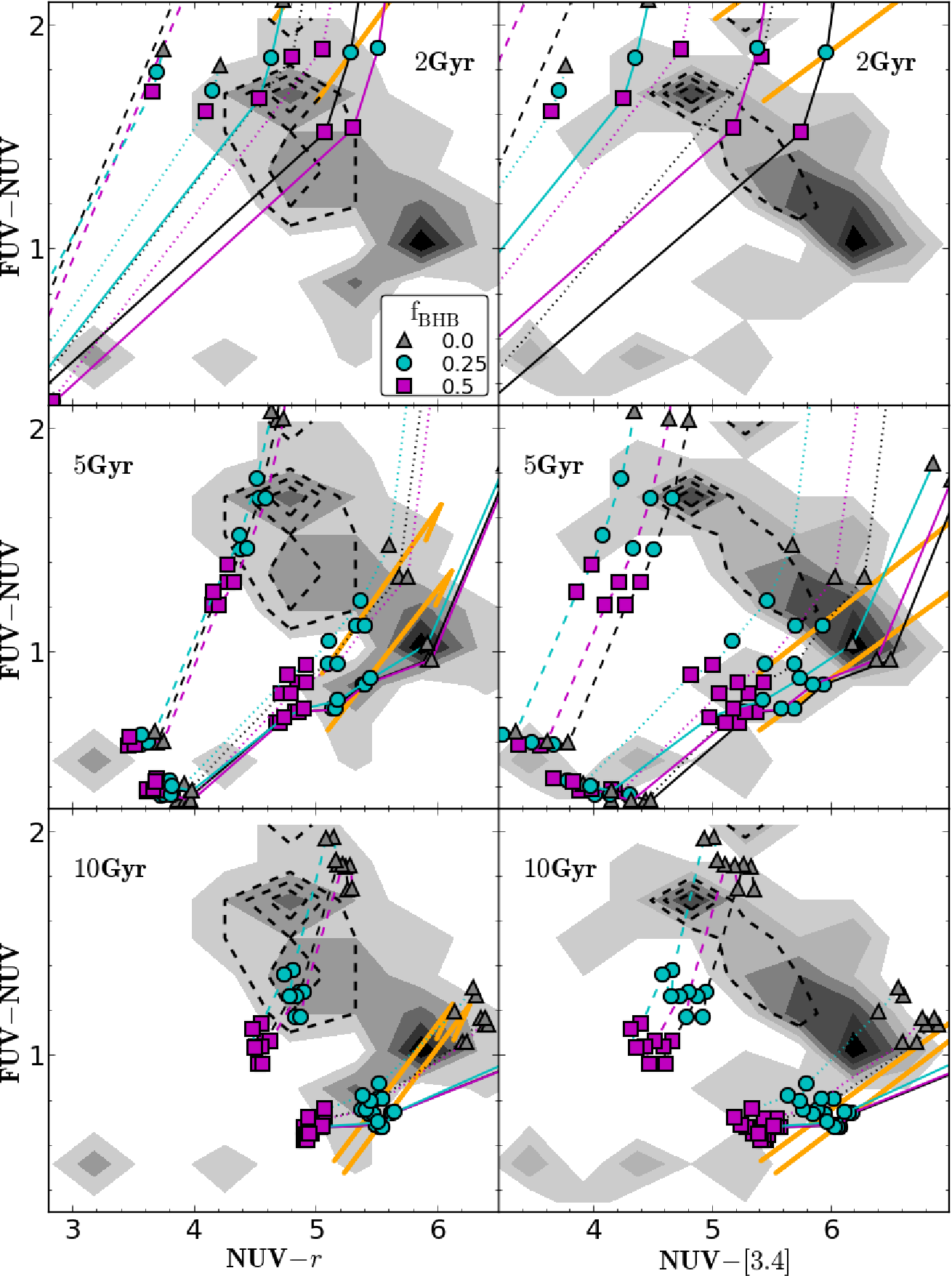}
\caption{FUV-NUV, \nuvr~ and NUV-[3.4] plots with the distribution of inner/outer colors, and isochrones at 2 (top), 5 (middle) and 10 (bottom) Gyr. The grey-filled contours include the inner and outer radii colors, while the outer regions are delineated with the dashed contour lines. The peaks of the two regions are clearly separated in color-space. The dashed, dotted and solid FSPS lines are Z $= 0.25$, 1, and 1.5 \zsun, respectively. The temperature increases are \delt $= 0$ (black), 0.2 (magenta), 0.4 (cyan). All lines include SFH e-folding times $\tau=0.2$, 0.6 and 1 Gyr. The BHB fraction at points along the isochrone tracks are marked by the symbols as labeled. We include dust vectors (orange) with an increasing dust parameter 0-1, using a \cite{cha00} dust model.
}\label{fig:isochrones}
\end{figure*}
}

\def\chisqage{
\begin{figure*}
  \epsscale{1}
  \plotone{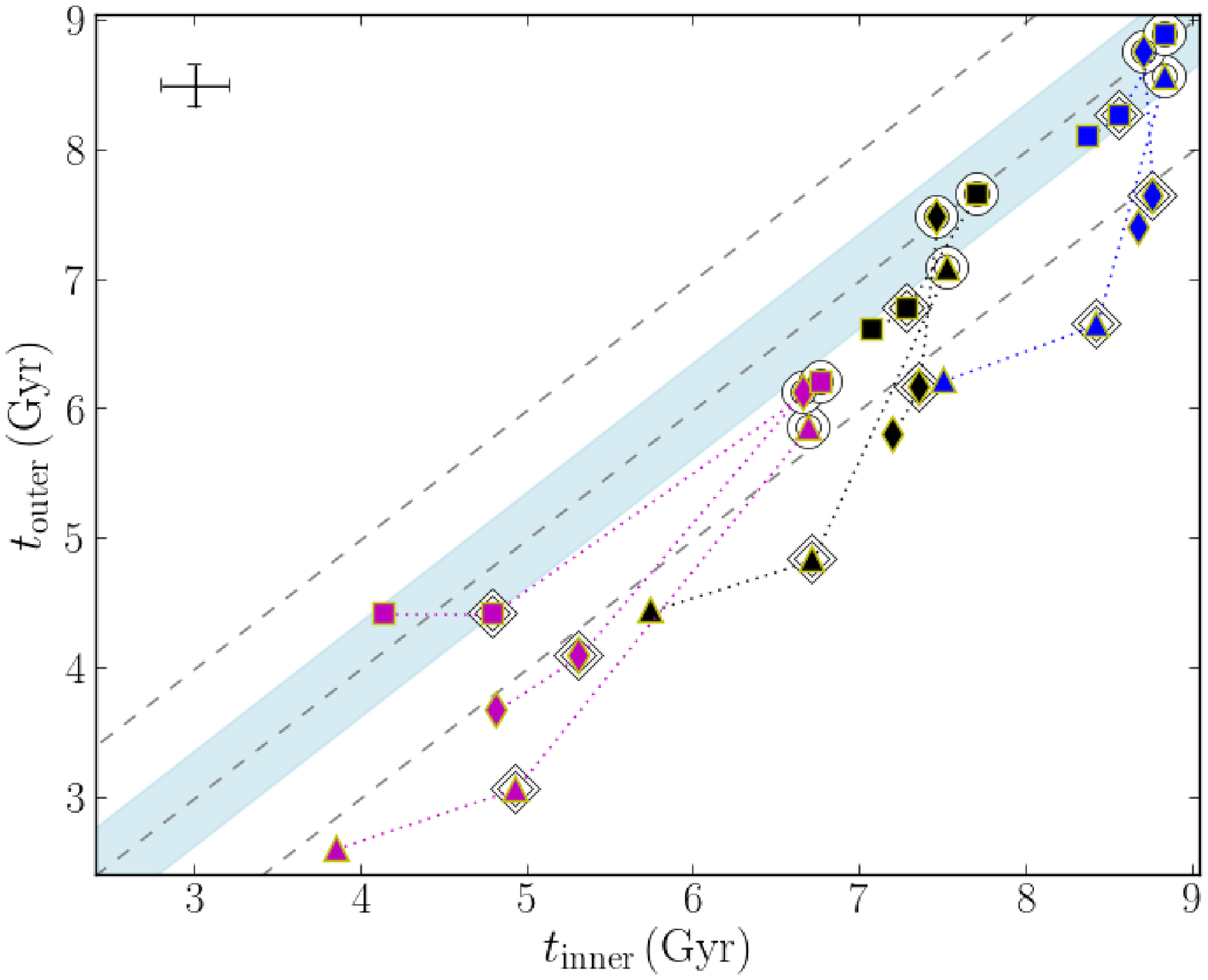}
  \caption{ Estimated ages in Gyr for \Rin~ and \Rout~ considering the different CSP parameter combinations, assuming homogeneity of the stellar populations in the inner and outer regions. The ages are combined weighted averages for the inner and outer regions for the 49 ETGs in our sample (one symbol is the average for all ETGs). The different triangles, diamonds and squares represent $f_{\rm BHB} = 0$, 0.25, and 0.5, respectively. The black, blue and magenta colors are $\tau=0.2$, 0.6, and 1, respectively. Symbols on double circles and double diamonds have Z = 0.25 and 1 \zsun, respectively, while all other symbols have Z=1.5 \zsun. The dotted lines trace metallicities along a fixed \fbhb~ and $\tau$ with metallicity. The shaded region highlights the equal age line within the average error. The average error is illustrated by the error bars in the upper left corner. We include equal age $\pm1$ Gyr dashed lines for gauging the age differences.
  }\label{fig:chisqage}
\end{figure*}
}

\def\ageZratios{
\begin{figure*}
  \epsscale{1}
  \plotone{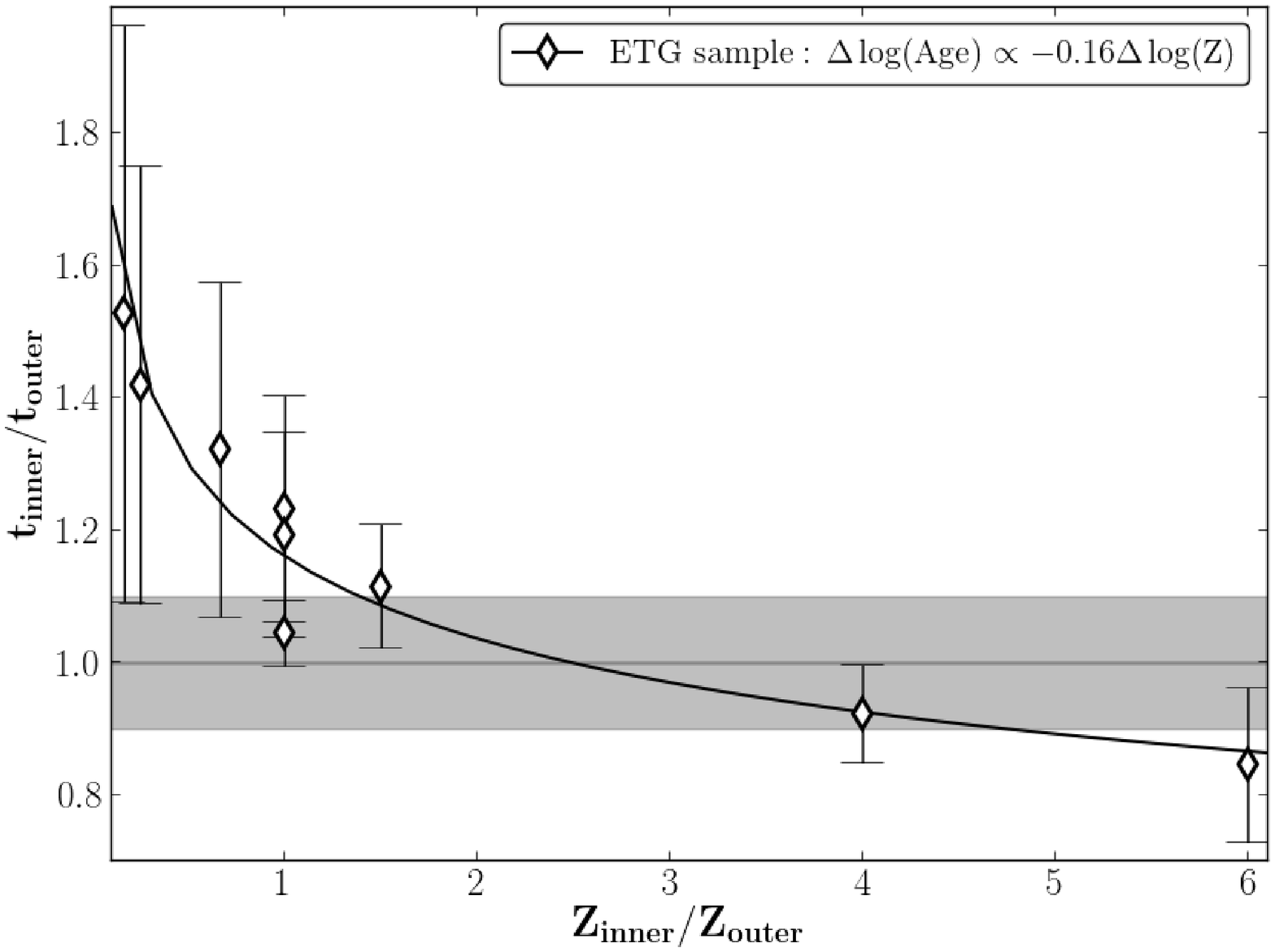}
  \caption{We plot the estimated age ratios for each metallicity combination between the inner and outer regions of the ETGs. The solid line is the fit of the ETG sample (diamonds). The error bars show the range of ages for the different BHB, $\tau$ combinations. The horizontal grey area shades the equal age line within a 10\% margin.
  }\label{fig:ageZratios}
\end{figure*}
}

\def\bhbclrdiffs{
\begin{figure*}
  \epsscale{1}
  \plotone{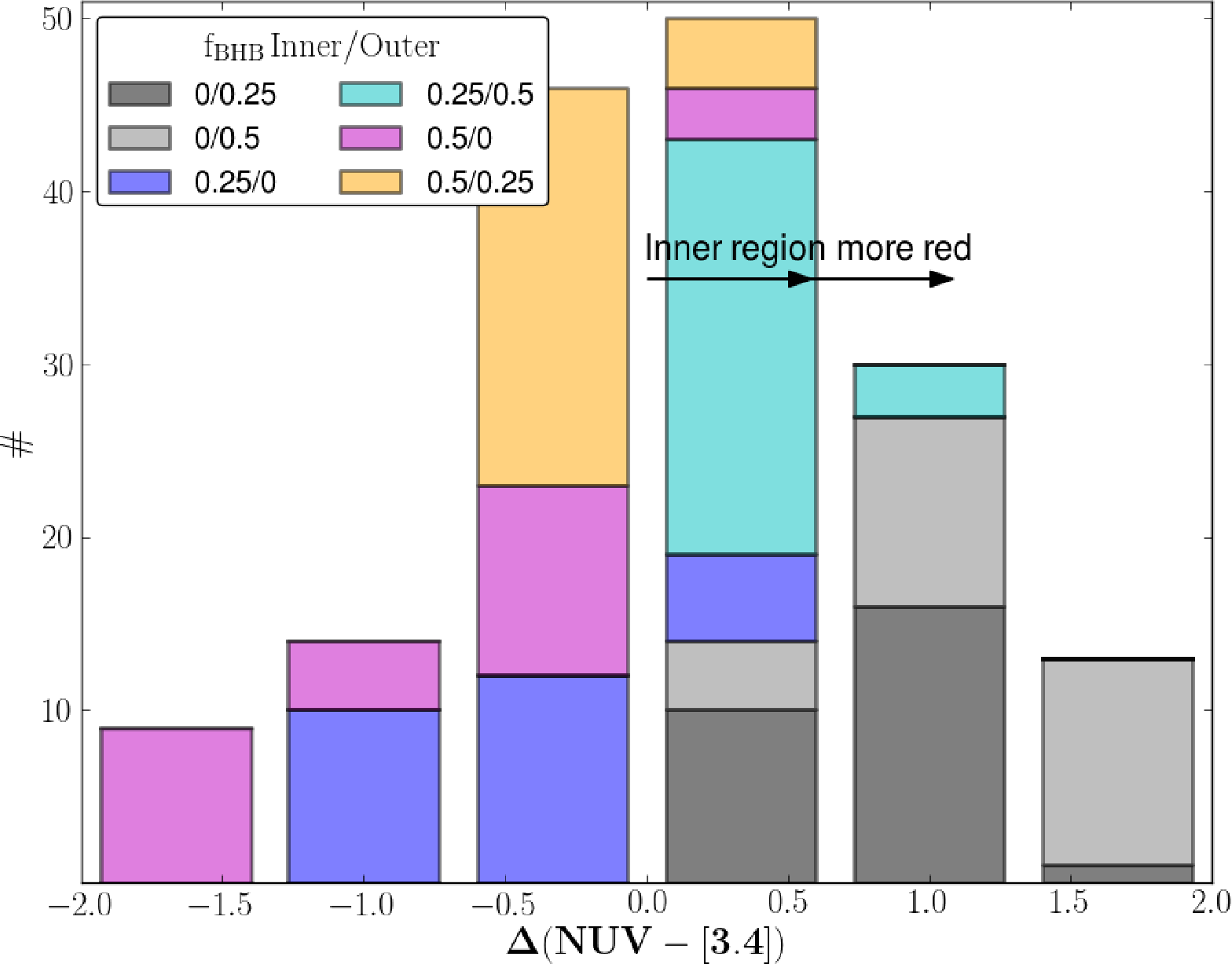}
  \caption{We show the stacked distributions of $\rm \Delta $(NUV-[3.4]) determined by imposing different BHB fractions at the inner and outer radii as labeled. For example, the dark grey labeled 0/0.25 refers to the color at \fbhb $= 0$ and 0.25 for \Rin~ and \Rout, respectively. The models with higher BHB fraction in the outer radii are more likely to have a 1 magnitude color difference (light blue, dark and light grey). 
  }\label{fig:bhbclrdiffs}
\end{figure*}
}

\def\massclr{
\begin{figure*}
\epsscale{1.5}
 \plotone{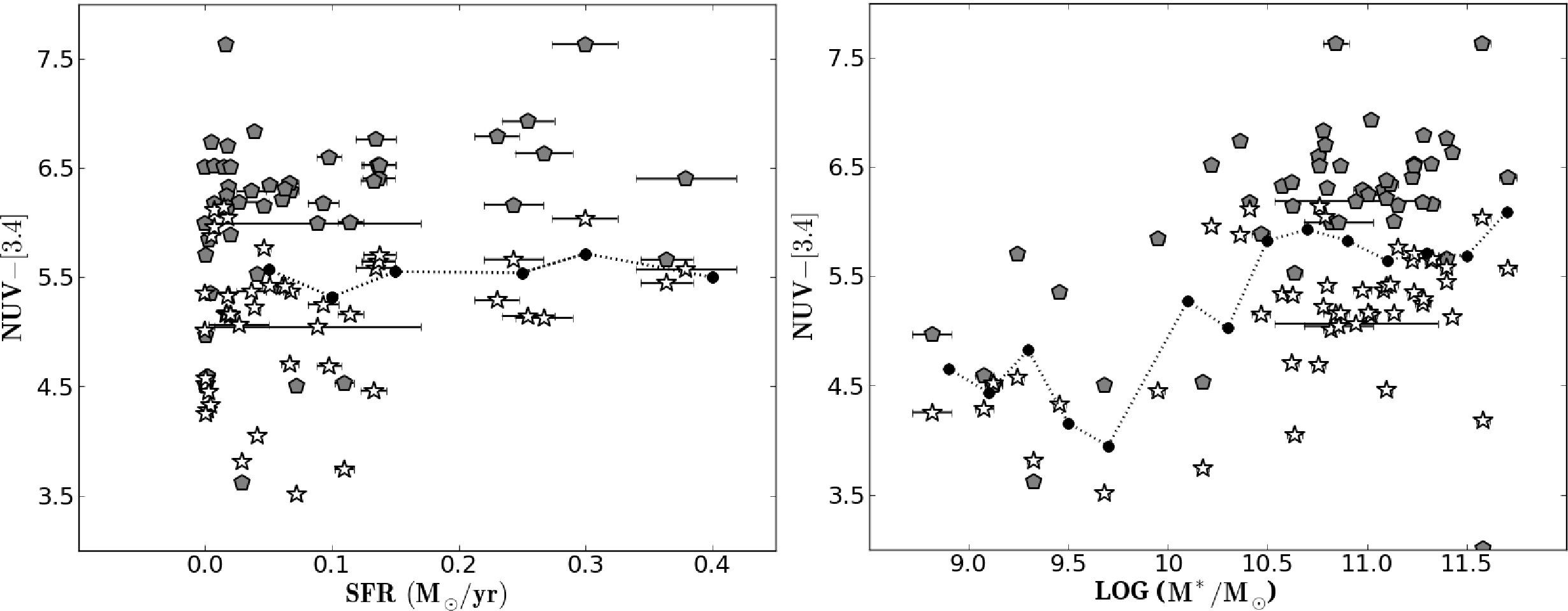}
\caption{NUV-[3.4] inner (grey pentagons) and outer (stars) region colors as a function of star formation rate (left) and stellar mass (right). The black circles and dotted line show the binned averages of the total flux NUV-[3.4] colors for SFR and $\rm M^*$ ($\rm \Delta SFR = 0.05 \, M_{\odot}\,yr^{-1}$ and $\rm \Delta \log (M^*/M_{\odot}) = 0.2$). As expected, the ETGs are quiescent, and there appears to be no correlation between the UV-mid-IR color and SFR. The average error for  NUV-[3.4] is 0.07 mag. The symbols are larger than the photometric errors.}
\label{fig:massclr}
\end{figure*}

}